\documentclass[longauth]{aa}  

\usepackage{graphicx}
\usepackage{txfonts}
\usepackage{lscape}
\usepackage[utf8]{inputenc}
\usepackage{threeparttable,pifont}
\usepackage{natbib}
\usepackage{color}
\usepackage{multicol}
\usepackage{amsmath,amssymb,url}
\usepackage{xcolor}
\usepackage{longtable,dcolumn,verbatim}
\bibpunct{(}{)}{;}{a}{}{,}
\usepackage{array} % fixes the corners
\usepackage[normalem]{ulem}

\usepackage{hyperref}
\begin{document}

   \title{Asteroid spin-states of a 4 Gyr collisional family. \thanks{This article is dedicated to the memory of Gianfranco Marcon, whose telescopes have contributed to the development of astronomy and to the observations collected for this study.}$^,$\thanks{Table B.5 is only available in electronic form at the CDS via anonymous ftp to cdsarc.u-strasbg.fr (130.79.128.5) or via \url{http://cdsweb.u-strasbg.fr/cgi-bin/qcat?J/A+A/}}}

   %\subtitle{}

   \author{D.~Athanasopoulos \inst{\ref{athens}}
        \and
        J.~Hanu{\v{s}} \inst{\ref{auuk}}%\fnmsep\thanks{Author who contributed the most.}
        \and
        C.~Avdellidou \inst{\ref{oca}}
        \and
        R.~Bonamico \inst{\ref{bsa}}
        \and
        M.~Delbo \inst{\ref{oca}}
        \and
        M.~Conjat \inst{\ref{oca}}
        \and
        A.~Ferrero\inst{\ref{ferrero}}
        \and
        K.~Gazeas \inst{\ref{athens}}
        \and
        J.P.~Rivet \inst{\ref{oca}}
        \and
        N.~Sioulas \inst{\ref{noak}}
        \and   
        G.~van Belle \inst{\ref{lowell}}
        \and
        P.~Antonini\inst{\ref{antonini}}
        \and
        M.~Audejean\inst{\ref{audejean}}
        \and
        R.~Behrend\inst{\ref{behrend}}
        \and
        L.~Bernasconi\inst{\ref{bernasconi}}
        \and
        J.W.~Brinsfield \inst{\ref{brinsfield}}
        \and
        S.~Brouillard\inst{\ref{aq}}
        \and
        L.~Brunetto\inst{\ref{brunetto}}
        \and
        M.~Fauvaud\inst{\ref{fauvaud}}
        \and
        S.~Fauvaud\inst{\ref{fauvaud}}
        \and
        R.~Gonz\'alez\inst{\ref{gonzales}}
        \and
        D.~Higgins\inst{\ref{higgins}}
        \and
        T.W.-S.~Holoien\inst{\ref{holoien1},\ref{holoien2}}
        \and
        G.~Kobber\inst{\ref{kobber}}
        \and
        R.A.~Koff\inst{\ref{koff}}
        \and
        A.~Kryszczynska\inst{\ref{amu}}
        \and
        F.~Livet\inst{\ref{livet}}
        \and
        A.~Marciniak\inst{\ref{amu}}
        \and
        J.~Oey\inst{\ref{oey}}
        \and
        O.~Pejcha \inst{\ref{utf}}
        \and
        J.J.~Rives\inst{\ref{aq}}
        \and
        R.~Roy\inst{\ref{roy}}
    %      \and
    %      K. Tsiganis \inst{8}
     }

   \institute{Section of Astrophysics, Astronomy and Mechanics,  Department of Physics, National and Kapodistrian  University  of Athens, Zografos GR 15784, Athens, Greece\label{athens}\\
        \email{dimathanaso@phys.uoa.gr}
        \and
    Charles University, Faculty of Mathematics and Physics, Institute of Astronomy, V Hole\v sovi\v ck\'ach 2, CZ-18000, Prague 8, Czech Republic\label{auuk}\\
        \email{josef.hanus@mff.cuni.cz}
        \and
    %Université C\^ote d’Azur, CNRS, OCA, LAGRANGE, France
    Université C\^ote d’Azur, Observatoire de la Côte d’Azur, CNRS, Laboratoire LAGRANGE, France\label{oca}
        \and
    BSA Osservatorio (K76), Strada Collarelle 53, 12038 Savigliano, Cuneo, Italy\label{bsa}
        \and
    Bigmuskie Observatory (B88), via Italo Aresca 12, 14047 Moberelli, Asti, Italy\label{ferrero}
        \and
    NOAK Observatory (L02), Stavraki Ioannina, Greece\label{noak}
    \and
    Lowell Observatory, 1400 West Mars Hill Road, Flagstaff, AZ 86001, USA\label{lowell}
    \and
    Observatoire des Hauts Patys, F-84410 B\'edoin, France\label{antonini}
        \and
    Observatoire de Chinon, Mairie de Chinon, 37500 Chinon, France\label{audejean}
        \and
   Geneva Observatory, CH-1290 Sauverny, Switzerland\label{behrend}
        \and
    Observatoire des Engarouines, 1606 chemin de Rigoy, F-84570 Malemort-du-Comtat, France\label{bernasconi}
    \and
    Via Capote Observatory, Thousand Oaks, CA 91320, USA\label{brinsfield}
    \and
    AstroQueyras, 05530 Saint-Véran, France\label{aq}
    \and
    Le Florian, Villa 4, 880 chemin de Ribac-Estagnol, F-06600 Antibes, France\label{brunetto}
     \and
    Observatoire du Bois de Bardon, 16110 Taponnat, France\label{fauvaud}
         \and
    Courbes de rotation d'ast\' ero\" ides et de com\` etes\label{gonzales}
        \and
    Hunters Hill Observatory, 7 Mawalan Street, Ngunnawal ACT 2913, Australia\label{higgins}
        \and
    NHFP Einstein Fellow\label{holoien1}
        \and
    The Observatories of the Carnegie Institution for Science, 813 Santa Barbara St., Pasadena, CA 91101, USA\label{holoien2}
        \and
    Courbes de rotation d'ast\' ero\" ides et de com\` etes\label{kobber}
        \and
    Antelope Hills Observatory, 980 Antelope DR W, Bennett, CO 80102 USA\label{koff}
        \and
    Astronomical Observatory Institute, Faculty of Physics, A. Mickiewicz University, S{\l}oneczna 36, 60-286 Pozna\'n, Poland\label{amu}
        \and
    Innstitut d’Astrophysique de Paris, 98 bis boulevard Arago, UMR 7095 CNRS et Sorbonne Universit\'es, 75014 Paris, France\label{livet}
    \and
    Blue Mountains Observatory, Leura, Australia\label{oey}
        \and
    Charles University, Faculty of Mathematics and Physics, Institute of Theoretical Physics, V~Hole{\v s}ovi{\v c}k{\'a}ch 2, 18000 Prague, Czech Republic\label{utf}
        \and
   Observatoire de Blauvac, 293 chemin de St Guillaume, F-84570 77 Blauvac, France\label{roy}
        %    Department of Physics, Aristotle University of Thessaloniki, GR 54124 Thessaloniki, Greece
 }
   \date{June 2022}

% \abstract{}{}{}{}{} 
% 5 {} token are mandatory
 
  \abstract
   {Families of asteroids generated by the collisional fragmentation of a common parent body have been identified using clustering methods of asteroids in their proper orbital element space. However, there is growing evidence that some of the real families are larger than the corresponding cluster of objects in orbital elements, and  there are families that escaped identification by clustering methods. An alternative method has been developed in order to identify collisional families from the correlation between the asteroid fragment sizes and their proper  semi-major axis distance from the family centre  (V-shape). This method has been shown to be effective in the cases of the very diffuse families that formed billions of years ago.}
   {Here we use multiple techniques  for observing asteroids to provide corroborating evidence that one of the groups of asteroids identified as a family from the correlation between size and proper semi-major axis of asteroids are real fragments of a common parent body, and thus form a collisional family. }
   {We obtained photometric observations of asteroids in order to construct their rotational light curves; we combine them with the literature light curves and sparse-in-time photometry; we input these data in the  light curve inversion methods, which allow us to determine a convex approximation to the 3D shape of the asteroids and their orientation in space, from which we extract the latitude (or obliquity) of the spin pole in order to assess whether an object is prograde or retrograde. We included in the analysis spin pole solutions already published in the literature aiming to increase the statistical significance of our results. The ultimate goal is to assess whether we find an excess of retrograde asteroids on the inward side of the V-shape of a 4 Gyr asteroid family identified via the  V-shape method. This excess of retrograde rotators is predicted by the theory of asteroid family evolution.}
   {We obtained the latitude of the spin poles for 55 asteroids claimed to belong to a 4 Gyr  collisional family of the inner main belt that consists of low-albedo asteroids. After  re-evaluating the albedo and spectroscopic information, we found that nine of these asteroids are interlopers in the 4 Gyr  family. Of the 46 remaining asteroids, 31 are found to be retrograde and 15 prograde. We also found that these     retrograde rotators have a very low probability (1.29\%) of being due to random sampling from an underlying uniform distribution of spin poles.}
   {Our results constitute  corroborating evidence that the asteroids identified as members of a 4 Gyr  collisional family have a common origin, thus strengthening their family membership.}


   \maketitle
%
%-------------------------------------------------------------------

\section{Introduction}
The study of asteroid families has been an active field of  research  since the discovery of the first groupings of asteroids in orbital element space \citep{hirayama1918}. As more asteroids were discovered, these initial groupings became more numerous, thus substantiating their significance. At the same time, more asteroid groupings (i.e. families) were discovered. Studies of the physical properties of asteroids  highlighted that the families were also homogenous in colour, albedo, and spectral properties, in general \citep[see][for a review]{Masiero2015}. This  corroborated the idea that these groups of asteroids in orbital element space were fragments of a common parent body. This the reason why they are also called collisional asteroid families \citep[see][for a review on the subject]{Nesvorny2015}.

The identification of the collisional families has been done using classical methods, such as the hierarchical clustering method \citep[HCM;][]{zappala1990asteroid, zappala1995asteroid}. Surveys of identification of asteroid collisional families have found that about one-third of the known asteroid population is associated with over 120 collisional families (see e.g. the Minor Planet Physical Properties Catalogue, MP3C) \footnote{\url{mp3c.oca.eu}}. However, it is well known that the large majority of asteroid family identification surveys are conservative in order to clearly identify the core of the family and keep a good separation between the nearby families as well \citep{Nesvorny2015}.

A very important question is  how many of the asteroids that are included in the background population  (i.e. that do not belong to families) of the main belt are instead collisional members that have not been associated with known families. There is evidence that this number is very large, implying  (i) that collisional families are larger than  is reported in our current catalogues \citep{Broz2013a, tsirvoulis2018, dermott2021dynamical} and (ii) that there are still undiscovered families. This problem derives from the difficulty in linking  the collisional members that currently reside far away in orbital elements to the family core \citep{milani2014}. It has also been shown that there is an unexpected lack of asteroid families that formed more than 2~Gyr years ago from a parent body larger than about 100~km \citep{Broz2013a}. Moreover, \citet{tsirvoulis2018} attempted to identify families using an aggressive version of the HCM, such that it was very generous in linking large number of asteroids to their respective families. These authors found evidence of undetected family members in the investigated region of the main belt from the study of the size frequency distribution of the non-family members. These observations also corroborate      hypotheses (i) and (ii).

As a collisional family ages, a non-gravitational force known as the Yarkovsky effect \citep[see e.g. the review in][]{Vokrouhlicky2015} pushes asteroids away from the centre of the family with a drift rate d$a$/d$t$ that is proportional to the inverse diameter (1/$D$), where $a$ is the asteroid orbital semi-major axis. Prograde rotating asteroids have d$a$/d$t$>0 and move to  a larger semi-major axis, while retrograde asteroids, with d$a$/d$t$<0, move to a smaller semi-major axis. This creates correlations of points in the (1/$D$ vs $a$) plane called V-shapes, because they resemble the letter V, whose slope ($K$) indicates the family age. As an asteroid family spreads in the orbital semi-major axis, family members cross orbital resonances with planets, which perturb their eccentricity and inclination. Hence, old families are less compact than younger ones in all three proper orbital element space. This makes old families (whose members have large orbital element spreading) more difficult to be identified by the HCM \citep[e.g.][]{bolin2017yarkovsky}, and cause families to overlap to each other. 

Based on the V-shape characteristic, a method was developed  to discover old and dispersed asteroid families  \citep{bolin2017yarkovsky,
Delbo2017}. This V-shape identification method has   already been used to discover five families of the inner main belt: the Eulalia and New Polana families \citep{Walsh2013}; a `primordial' family with a nominal age of $4.0^{+1.7}_{-1.1}$~Gyr, but that could be as old as the Solar System; the Athor family, $3.0^{+0.5}_{-0.4}$~Gyr; and the Zita family,  $5.0^{+1.6}_{-1.3}$~Gyr \citep{Delbo2017, delbo2019ancient}. 

Although the V-shape family identification method is indeed a powerful tool, its efficiency   decreases with increasing family age \citep{deienno2021efficiency} because the most dispersed family members can no longer be distinguished  from the background, which might also consist of old and dispersed overlapping families. 
Given the current limitations of the detection methods, we are looking  for independent ways to confirm family members and thus the borders of the families identified on the basis of their V-shapes. One of these independent methods is to check the spin state of the family members. According to the  theory of the Yarkovsky effect, it is expected that most of the asteroids on the inward side (the left side of the V) of the family are retrograde, and prograde on the outward side (the right side of the V). This has been shown  in notable HCM asteroid families \citep{Hanus2013c,Hanus2018a}.

One of the most effective techniques to identify the spin state of asteroids is the inversion of their photometric light curves \citep{kaasalainen2001optimization1, kaasalainen2001Optimization2}. This method requires the acquisition of large datasets of photometric measurements, where most of them are currently retrieved from all-sky surveys. 
However, these data are sparse in time by nature and often do not allow the unambiguous determination of asteroid rotational periods due to their low photometric accuracy of typically 0.1 magnitude. Often the  addition of classical dense-in-time optical light curves to the sparse dataset leads to the removal of such ambiguity.

In this work we study the spin states of asteroids that belong to the innermost border of the inward side of a primordial family of the inner main belt that was reported by \cite{Delbo2017}. This primordial family has orbital elements ($a,e,i$) roughly ($2.26,0.14,5.75^\circ$) for the inward wing as only this has been identified and is suspected to contain low-albedo C-complex asteroids. For this, we ran our own observing programme (called {\it Ancient Asteroids}) and obtained dense-in-time light curves, which we combined with existing sparse-in-time photometric data.

In Sect.~\ref{sec:data} we present the optical datasets used in this work and present our observing campaign. In Sects.~\ref{sec:LI}~and~\ref{sec:spinpoles} we describe the light curve inversion method used for the analysis of the optical data and for the determination of asteroidal physical properties. In Sect.~\ref{sec:results} we present our results on the spin poles of our dataset indicating also the potential asteroid interlopers of the primordial family, while in Sect.~\ref{sec:discussion} we discuss the results.

%--------------------------------------------------------------------
%\section{Observations and Data Acquisition}
\section{Datasets}\label{sec:data}
In order to verify membership in the primordial family, we focus our analysis on the asteroids that are located between the inward border of the primordial family and the respective border of the Polana family (see Fig.~\ref{fig:ObsByUs}). The latter, which is also called New Polana after its reassessment by \cite{Walsh2013}, is another large but younger family in the inner main belt of similar carbonaceous composition and low albedo. We use the  V-shapes of \cite{Walsh2013} and \cite{Delbo2017} to distinguish between the two families. This group of objects should belong either to the vast and extensive primordial family or to the unidentified background population. 
In order to study the spin poles of the primordial family we combined a great deal  of data, including  asteroid light curves that we collected from the databases,  sparse photometric data obtained from different surveys,  existing complete or incomplete shape models, and  our own photometric observations. 

%-------------------------------------------------------------------
\subsection{Currently available asteroid models} 
The vast majority of asteroid shape models have been produced by the light curve inversion method \citep{kaasalainen2001optimization1, kaasalainen2001Optimization2}. The Database of Asteroid Models from Inversion Techniques (DAMIT)\footnote{\url{https://astro.troja.mff.cuni.cz/projects/damit/}} contains $\sim$6\,000 asteroid models for $\sim$3\,460 asteroids (March 2022) that are publicly available \citep{durech2010DAMIT}. Shape models for 15 asteroids from our list are already included in DAMIT (see Table~\ref{tab:LitPoles}). Furthermore, we also considered the  partial models published in \citet{Durech2020}. For these models only the sidereal rotation period,  the ecliptic latitude of the spin axis, and its range are reported. This information is sufficient for our purposes, as it often allows us to securely decide whether the asteroid is a prograde or a retrograde rotator.

%-------------------------------------------------------------------
\subsection{Archival photometric data}\label{sec:archive}
Photometric data of asteroids are scattered throughout various public databases or were provided to us directly by the observers. During the past two decades a large internal database of optical light curves  maintained at the Institute of Astronomy of Charles University has been routinely used for shape  modelling. A large amount of data was obtained from the Asteroid light curve Data Exchange Format (ALCDEF) database\footnote{\url{https://alcdef.org/}} \citep{stephens2018ALCDEF} (147 dense photometric light curves that exist for 19 of our asteroid targets). Additional dense light curves were downloaded from the Asteroid Photometric Catalogue \citep[APC;][]{lagerkvist2011asteroid} and the Courbes de rotation d'ast\' ero\" ides et de com\` etes database (CdR\footnote{\url{https://obswww.unige.ch/~behrend/page3cou.html}}) or were provided directly by the observers (see Table \ref{tab:ObsByUs}).

The sparse-in-time data that come from various sky-surveys were obtained from corresponding databases and archives connected to the publications. We used data from the US Naval Observatory in Flagstaff (USNO-Flagstaff, IAU code 689), the Catalina Sky Survey \citep[CSS, IAU code 703;][]{Larson2003}, Gaia Data Release 2 \citep[GaiaDR2;][]{Spoto2018}, the All-Sky Automated Survey for Supernovae \citep[ASAS-SN;][]{Shappee2014,Kochanek2017,Hanus2021}, the Asteroid Terrestrial-impact Last Alert System \citep[ATLAS;][]{tonry2018, Durech2020}, the Zwicky Transient Facility \citep[ZTF, IAU code I41;][]{bellm2019}, the Palomar Transient Factory Survey \citep[PTF;][]{Chang2015, Waszczak2015}, and TESS \citep{Ricker2015, Pal2020}. Tables~\ref{tab:NewModels} and \ref{tab:RevisedModels} summarise the typical number of measurements from these surveys that were available for our targets. In general, data from USNO-Flagstaff, GaiaDR2, ZTF, TESS, and PTF are rather limited for our targets, while hundreds of individual measurements are available from  the CSS, ASAS-SN, and ATLAS surveys. The CSS, USNO-Flagstaff, and ZTF data were obtained through the AstDys-2 database\footnote{\url{https://newton.spacedys.com/astdys/}}. %Data from ASAS-SN ($V$-band), GaiaDR2, TESS, and PTF are parts of the corresponding publications. %Data from the ATLAS survey were obtained by \citet{Durech2020}.

\subsection{ASAS-SN $g$-band data}
As discussed in Sect.~\ref{sec:archive}, we use $V$-band sparse data from the ASAS-SN survey through the catalogue of \citet{Hanus2021}. However, since 2018 ASAS-SN has used  the SLOAN $g$ filter and significantly expanded to more telescope units and sites; we utilised these data in our work as well. We accessed and processed the $g$-band data following the same procedure as in \citet{Hanus2021}. The time coverage is already more than three years. Moreover, the $g$-band limiting magnitude is larger by about one magnitude than that of the $V$-band. Therefore, the $g$-band dataset is often comparable in terms of the number of measurements to the $V$-band dataset for brighter objects, and outperforms the $V$-band dataset for fainter objects. Both filters are treated independently in the shape modelling.

%-------------------------------------------------------------------
\subsection{Ancient Asteroids: An international observing campaign}
In order to enlarge our input dataset used for the shape modelling, which would potentially lead to new and improved shape solutions, we performed additional ground-based photometric observations. An international observing campaign has been initiated in the framework of our international initiative called {\it Ancient Asteroids}\footnote{\url{http://users.uoa.gr/~kgaze/ancient_asteroids.html}} in order to collect dense photometric data for asteroids that belong to the oldest asteroid families \citep{athanasopoulos2021ancient}. {\it Ancient Asteroids} establishes a network of astronomers, currently from four countries, who follow a common observing plan.
In the following we present the observing facilities and their corresponding equipment that participated in this work and provided data for 35 asteroids in our dataset. 

%-------------------------------------------------------------------
\subsubsection*{Bonamico Star Adventure Astronomical Observatory} %(K76)
The Bonamico Star Adventure (BSA) is an amateur observatory located in Savigliano, Italy. For this
study BSA used  a robotic 0.3~m (f/8) Ritchey-Chretien telescope and an open-filter SBIG ST-9 XME CCD detector.
 %\footnote{\url{http://observaorioastronomicobsa.it}}

%-------------------------------------------------------------------
\subsubsection*{Lowell Observatory}
Lowell is located in Arizona, United States, and for this project it operated two robotic telescopes:  the Titan Monitor Telescope (TiMo), a 20" PlaneWave CDK20 telescope equipped with a Moravian instruments G3-6300 CCD detector, and  the  1~m PlaneWave (PW1) telescope equipped with a Finger Lakes Instruments ML-16803 CCD imager. TiMo is located on Lowell's main Mars Hill campus (IAU Code 690), whereas the PW1 is on Lowell's Anderson Mesa campus (IAU Code 688). All the observations were performed in the  Sloan r' filter.
%There is a 20-position filter wheel on TiMo with a set of Sloan ugriz filters, as well as 12 narrowband filters between 340 and 880~nm.
%Both telescopes operate robotically using ACP DC-3 software. 
%\footnote{\url{https://lowell.edu}}
%-------------------------------------------------------------------
\subsubsection*{Bigmuskie Observatory} %(B88) 
Bigmuskie is an amateur observatory located in Mombercelli-Asti, Italy. Bigmuskie utilises a 0.4~m (f/8.25) Ritchey- Chr\'{e}tien telescope. The observations were performed unfiltered by using a Moravian G3-1000 CCD detector.
%-------------------------------------------------------------------

\subsubsection*{Observatoire de la Côte d'Azur}
The observations were performed at two stations, which belong to the Observatoire de la Côte d'Azur (OCA) in France. The first station is the C2PU facility in Calern, at an altitude of a 1300~m. C2PU operated the 1.04~m Cassegrain telescope (known as Omicron@C2PU) with an f/3.2 parabolic, prime focus and with a three-lens Wynne coma corrector using an unfiltered QHY600 CMOS camera \citep{bendjoya2012C2PU}. The second station is located on Mont Gros, the historical site of the OCA, on the east-side hills of the town of Nice, France. Mont Gros station operated the 0.4~m (f/5) diameter telescope (called Schaumasse) equipped with with a QSI 583ws CCD camera.
%\footnote{\url{https://www.oca.eu/en/home-oca-en}}
%One of them (the westmost telescope, named "Omicron@C2PU") can be fully remote-operated and has been used for this asteroids photometry campaign.
%-------------------------------------------------------------------

\subsubsection*{University of Athens Observatory }
The University of Athens Observatory (UOAO) belongs to the National and Kapodistrian University of Athens in Greece, which utilises a robotic 0.4~m (f/8) Cassegrain telescope equipped with an SBIG~ST-10~XME CCD detector \citep{Gazeas2016robotic}. All the observations were performed unfiltered. 

%-------------------------------------------------------------------
\subsubsection*{Helmos Observatory}
Helmos observatory is operated by the National Observatory of Athens and is located on Mount Helmos (Aroania) in Greece, at an altitude of 2340~m. It utilises a robotic 2.3~m (f/8) Ritchey-Chr\'{e}tien telescope (called Aristarchos) \citep{goudis2010aristarchos}. All the observations were performed unfiltered by using the Princeton Instruments VersArray 2048B LN CCD camera. 

%-------------------------------------------------------------------
\subsubsection*{NOAK Observatory} %(L02)
The NOAK observatory is located in the city of Ioannina, Greece. It utilises a 0.25~m (f/4.7) robotic Newtonian telescope. All the observations were performed unfiltered by using an ATIK 460EXM CCD camera.

\subsubsection*{BlueEye 600 Observatory}
The BlueEye 600 robotic observatory (BE600) is operated by the Astronomical Institute of the Charles University and is located in Ond\v{r}ejov, Czech Republic. It utilises a 60~cm Ritchey-Chr\'{e}tien telescope (Officina Stellare). All the observations were performed by Martin Lehk\'y\footnote{Deceased November 18, 2020.} utilising the standard Johnson R filter and the E2V42–40 CCD camera \citep{durech2018shape}.

\subsubsection*{Pic de Ch\^ateau-Renard Observatory }
The  Pic de Ch\^ateau-Renard Observatory (ChR) is located at an altitude of 2936 m in Saint-V\'eran in the  French Alps. The  facility is operated by the Paris-Meudon Observatory and AstroQueyras, an amateur association. Observations were performed unfiltered by using a 0.5 m (f/8) Ritchey-
Chrétien telescope equipped with a SBIG STX 16803 camera.

\subsubsection*{Observatoire du Bois de Bardon}
The Observatoire du Bois de Bardon (OBdB) is an amateur observatory located in Taponnat, France. OBdB used a 0.28~m (f/3) Schmidt-Cassegrain telescope equipped with a SBIG ST-402 ME CCD camera. Observations were performed with an r' (Sloan) filter.

\subsubsection*{Blue Mountains Observatory}
The Blue Mountains Observatory is located at an altitude of 900~m in Leura, Australia. The photometric observations were done with a classical Celestron Schmidt Cassegrain telescope, 0.35~m in diameter operating at f/5. All images were taken unfiltered using a SBIG CCD camera ST8-XME at bin 1x1.

%-------------------------------------------------------------------
\section{Photometric reduction}\label{sec:LI}

The photometric datasets used in this work include both dense photometric data from ground-based facilities (retrieved from the literature or from our observing campaign), as well as sparse data from several sky surveys and space missions, as described above. These two different datasets require different analysis techniques. 

%-------------------------------------------------------------------
\subsection{Observations in the  Ancient Asteroids programme} \label{subsect:AAobs}

Our observations were performed mainly in clear filter in order to increase the signal-to-noise ratio in our light curves, while keeping the exposure time as short as possible and increasing the sampling frequency. The exposure time   varied between 30~s and 240~s, depending on the brightness of the target, telescope aperture, and observing conditions. 

The collected data from all the observatories were reduced, following the standard image processing procedure of calibration and aperture photometry \citep[e.g.][]{massey1997user, gallaway2020high}. The calibration was performed for all light frames in three steps:    bias subtraction,  dark subtraction, and   flat-field correction. Aperture photometry is a quite simple technique and most applicable to stellar fields that are relatively sparse. This procedure was compiled by utilising \textit{AIP4Win} software \citep{berry2005handbook} for the fields observed by the  C2PU, UOAO, Helmos, and NOAK observatories; IRIS software \footnote{\url{http://www.astrosurf.com/buil/iris-software.html}} for the images performed by the OBdB and ChR observatories; and \textit{MPO Canopus} software \citep{warner2015MPO} for the remaining images (see Table~B.5).

The differential photometry was performed either with five bright field stars or by estimating an artificial comparison star, following the methodology presented by \citet{broeg2005new}. The resulting measurements were provided in differential magnitudes with a photometric accuracy of 0.02--0.1~mag. In the case of   OBdB and ChR, the differential photometric data were performed by estimating an artificial comparison star following the methodology described by \citet{Fauvaud2013,Fauvaud2014}. 

We used the sigma-clipping method \citep[see][]{gallaway2020high} to remove the prominent outliers in our measurements, which were usually caused by cosmic rays or satellites passing through the field. In the case where an asteroid was passing near a field star within a range of an aperture size (typically of the order of 5-7 arcseconds) we trimmed the light curve and we kept only the `clear' parts.

We used the Pogson equation ($m=-2.5log(F)+c$) to transform the differential magnitude ($m$) to relative flux ($F$). The relative flux values were normalised by defining the average flux of each light curve as one. For all the epochs of the observed light curves we performed the light travel time correction (from the asteroid to the observer) and computed the ecliptic Cartesian coordinates ($x,y,z$) of the Sun and of the Earth, with the asteroid as the reference point, in [au] via the Miriade service \citep{berthier2009miriade}. This format is required by the convex inversion (CI) method that we used for the shape modelling, as described in Sect.~\ref{sec:spinpoles}.

\subsection{Adopted data}
The dense photometric data from databases such as the  Asteroid Lightcurve Data Exchange Format database (ALCDEF) or APC are in magnitude values, so we converted them to relative fluxes by following the   procedure   described in Sect.~\ref{subsect:AAobs}.

In addition to the  dense photometric data, we included    sparse-in-time photometric measurements from various sources as they proved to be useful in constraining the asteroid models, despite their usually low photometric accuracy of $\sim$0.1~mag \citep{Durech2009,Hanus2011, Hanus2013a, Durech2016}. In order to use the sparse data for the shape modelling by the CI method, we processed them following the procedure of \citet{Hanus2011}. For more details, we refer to the most recent description of the procedure applied to ASAS-SN data by \citet{Hanus2021}. All individual measurements within each sparse dataset (i.e. specific survey and photometric filter) are internally calibrated; therefore, we process each dataset separately. First, the sparse data are usually available in magnitude values, which we transform into fluxes utilising the Pogson equation and, for convenience, setting the zero magnitude to 15. We then apply the light travel time correction to each epoch. Next we normalise the fluxes to a referenced one astronomical unit distance of the asteroid to the Earth and the Sun. The final steps were sigma-clipping to reject the outliers and estimating the  relative weights of each sparse dataset with respect to the dense data \citep[see][]{Hanus2021}.

%-------------------------------------------------------------------

\section{Determination of the spin poles} \label{sec:spinpoles}

We used the CI method developed by \citet{kaasalainen2001optimization1,kaasalainen2001Optimization2}. This gradient-based inversion technique is based on shape model parametrisation by a set of facets and their normal vectors and their optimisation such that they fit to the observed light curves. Assuming a convex shape representation of the asteroid shape, the inversion problem is unique. However, adding the rotation state (i.e. sidereal rotation period and spin axis orientation) as additional free parameters, we lose the uniqueness of the solution, and the parameter space becomes full of local minima. We have to search the parameter space on a grid of input parameters and find the local minimum that corresponds to the global minimum, and thus the correct set of searched parameters. The production of the model light curve, that is compared by the method to the observed light curves, is performed by using an empirical light-scattering model, which  is a combination of single Lommel-Seeliger and multiple Lambert scattering models \citep{kaasalainen2002steroid}. To date, the CI has been used to derive asteroid models for more than 3\,460 asteroids that are stored in the DAMIT database.

We assume that the shape effects in the light curves are independent of the  photometric filters used while covering the reflected-dominated spectral range. Therefore, we can treat all dense light curves as relative (i.e. normalised to unity). Although each sparse dataset is, in principle, internally calibrated in a different photometric system, we also use the sparse data  as normalised to unity. The only caveat is that we assume that the phase function is the same in each photometric system. This is not fully correct; for example,  \citet{Durech2020} found  statistically significant differences in the ATLAS data taken in the  c and o filters, in accordance with the phase reddening effect \citep{Millis1976,Lumme1981b}. However, for our purposes it is sufficient to have a single phase function for all sparse datasets.

We combined all the available datasets and applied   CI to them. We weighted individual light curves and sparse datasets based on their accuracy (expected rms). The individual weights $w_i$ were normalised such that $\sum(w_i) = N$, where $N$ is the number of dense light curves plus the number of sparse datasets. The process that we followed is described in Hanu\v s et al. (in prep.).

\subsection{Rotation period}

For asteroids with previously known rotation periods we searched for the best-fit model with the period parameter varying between boundaries defined by 5\%, 10\%, and 20\% of the previously reported period in the LightCurve DataBase \citep[LCDB;][]{Warner2009}, depending on the reliability flag provided for each period estimate 3 and 3--, 2+, 2. Rotation periods for asteroids with other reliability flags were considered  unknown.

Fourier-based algorithms are unable to efficiently estimate the rotation period of asteroids in cases of only sparse photometric data availability or extreme slow rotators, where their rotation period far exceeds the night duration. 
In these cases we performed a dense scanning in the rotation period parameter space from 2~h, which has been observed as an approximate lower limit of asteroids (>150~m in size)  by \cite{pravec2002asteroid}, up to 5\,000~h, which was motivated by the recent discovery of superslow rotating asteroids by \cite{erasmus2021discovery}.

It should be noted  that the period search is actually a full shape and rotation state optimisation by the CI. However, this procedure was performed on a grid of pole orientations limited to only ten values evenly distributed on a sphere and rather rough shape model resolution. We only recorded the rotation period and the  best-fit rms (and $\chi^2$) value within the grid of pole directions. We considered the best-fitting period searched for on a selected period interval as unique if its $\chi^2_{\mathrm{min}}$ is the only solution below the threshold defined as

\begin{equation}\label{eq:chi2limit}
\chi^2_{\mathrm{tr}} = \left(1+0.5\sqrt{\frac{2}{\nu}}\right)\chi^2_{\mathrm{min}},
\end{equation}
where $\nu$ corresponds to the number of degrees of freedom (number of observations minus the number of free parameters).

 Interestingly, the difference between two local minima in the period parameter space $\Delta P$ is dependent only on the time span of the data $T$ and the sidereal rotation period $P$ itself 
\begin{equation}\label{eq:deltaP}
    \frac{\Delta P}{P} = \frac{1}{2}\frac{P}{T},
\end{equation}
and thus, for instance, independent of the shape resolution. We sample the period with a step of 0.5$\Delta P$ that we recompute every step. We note that $\sim1/20$ of $\Delta P$ is a reliable uncertainty for the derived rotation period as it corresponds to a rotation phase offset of about 10$^\circ$, which is a typical uncertainty on the pole direction.

\subsection{Spin pole and shape}
After the determination of the rotation period, we applied the CI method on a much denser grid of initial spin directions (ecliptic longitude $\lambda$ and latitude $\beta$) and a higher shape resolution\footnote{For the search for the rotation period we fit 7$^2$ coefficients of the spherical harmonic expansion and start with ten different pole directions uniformly distributed on the sphere. The denser grid utilises 9$^2$ coefficients of the expansion and 48 input pole directions. The typical polyhedron representing the final shape solution contains about 1,000 vertices and 2,000 facets.}. The best-fitting shape and spin pole solution is considered unique if it fulfils the condition of Eq.~\ref{eq:chi2limit}. Most of the cases presented two symmetrical solutions with respect to ecliptic longitude ($\lambda \pm$ 180$^{\circ}$), the so-called pole ambiguity \citep{Kaasalainen2006}. 

If photometric data are rich enough in observing geometries and have reasonable accuracy relative to the amplitude of the asteroid light curve, the unique solution can be often derived. Sometimes, however, the photometric data are insufficient to derive a unique solution, but they allow us to derive a unique period and three or four pole solutions that have similar values of the ecliptic latitude $\beta$. These so-called partial models are still useful, especially for our study, as it is possible to decide whether the asteroid is a prograde or a retrograde rotator. Therefore, we made an effort to identify such cases here together with the unique solutions.

%-------------------------------------------------------------------
%-------------------------------------------------------------------
\section{Results}\label{sec:results}
\begin{figure*}[!htbp]
    \centering
    \includegraphics[width=\linewidth]{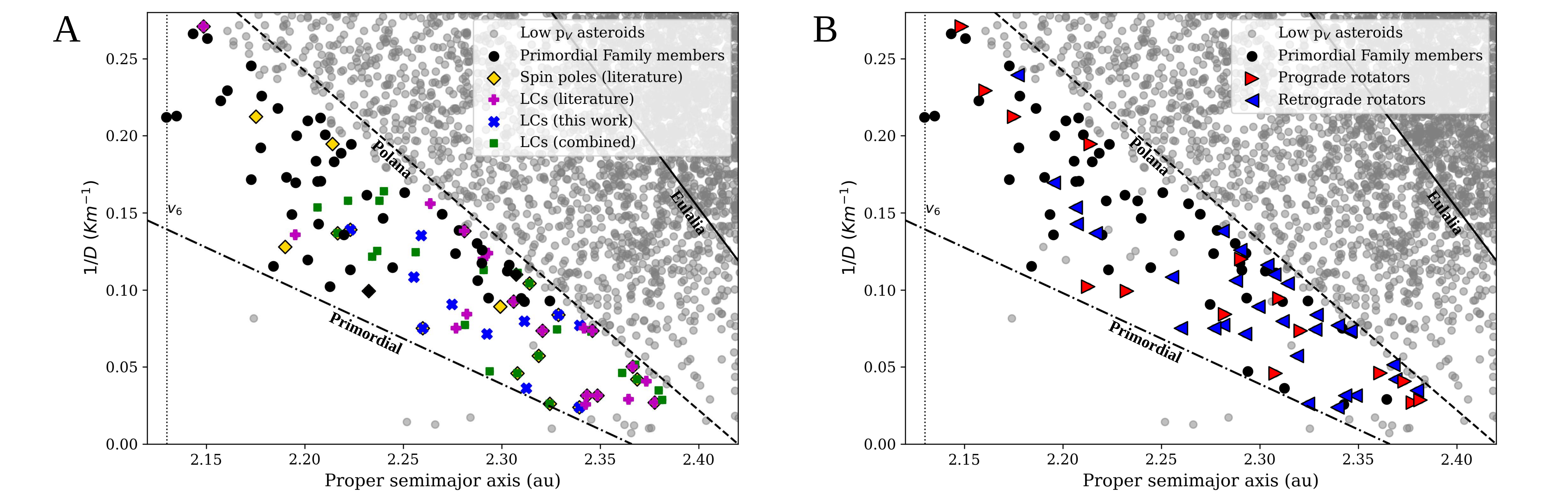}
    \caption{Primordial family members presented in proper semi-major axis vs inverse diameter plane, along with the low-albedo asteroids located in the innermost region of the main belt. Panel A: The yellow diamonds indicate  members with known spin pole from the literature. The plus signs, crossed, and squares   are members for which there are dense light curves only from literature, from this work, and from both sources, respectively (see  details in Table \ref{tab:ObsByUs}). Panel B: Left side of the V-shape of the primordial family. The prograde asteroids are shown in red and  the retrograde asteroids in blue.}
    \label{fig:ObsByUs}
\end{figure*}

\begin{figure}[!ht]
    \centering
    \includegraphics[width=\columnwidth]{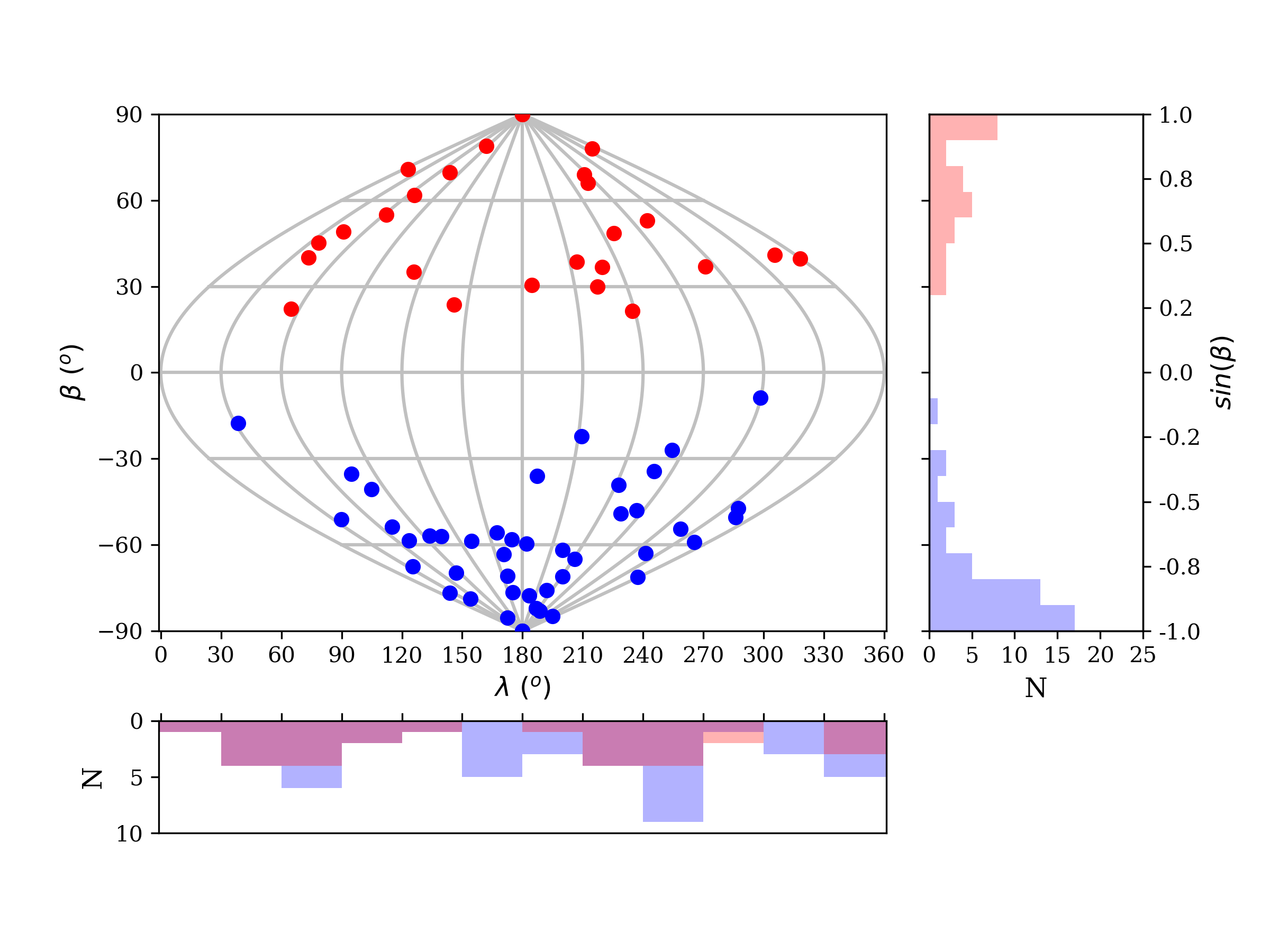}
    \caption{Distribution of the complete spin pole solutions for the primordial family members. The main plot is a sinusoidal equal-area cartographic representation, where the vertical grey lines define the longitude ($\lambda$) and the horizontal curves define the latitude ($\beta$). The right histogram represents the latitude ($\beta$) of prograde (red) and  retrograde (blue) rotators. The bottom histogram represents the longitude ($\lambda$) of prograde (red) and  retrograde (blue) rotators. The specific values for each asteroid are shown in Tables~\ref{tab:NewModels}~and~\ref{tab:RevisedModels}.}
    \label{fig:spd}
\end{figure}

\begin{figure}[!ht]
    \centering
    \includegraphics[width=\columnwidth]{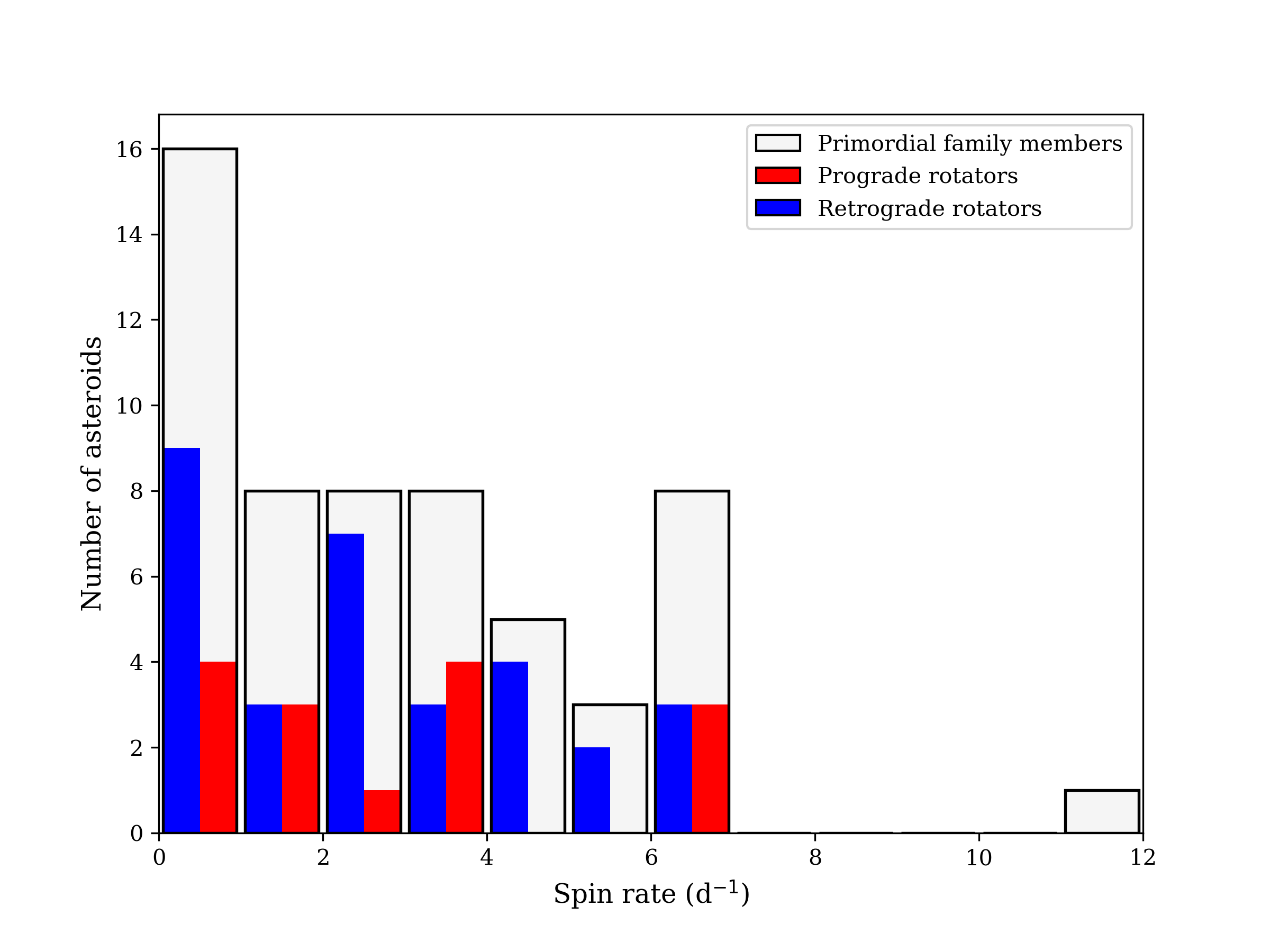}
    \caption{Histogram of spin rate for members of the primordial asteroid family. The prograde rotators are in  red  and the retrograde in blue. The light grey bars represent the primordial family members whose rotational period is known either from this study or the literature.}
    \label{fig:Phist}
\end{figure}
%-------------------------------------------------------------------
\subsection{New period estimates}

We derived improved values of the sidereal rotation period for 29 asteroids, as presented in Tables~\ref{tab:NewModels}~and~\ref{tab:RevisedModels} (see also Fig.~\ref{fig:periodograms} for an example of typical periodograms). All these period values are consistent with the synodic periods reported in the LCDB database\footnote{The only exception is asteroid (1159)~Granada with the reported LCDB period of 77.28~h. However, this period adopted from the CdR database is preliminary and inconsistent with the period reported, for example, by \citet{Waszczak2015}, which is  similar to the value we derived.}. The revised asteroid models have almost the same period as the previous solutions. 
Moreover, we measured rotation periods for seven asteroids for the first time. These periods were derived using sparse photometric data. Of these asteroids one is a super-slow rotator with $P = 3253.5$~h and one a slow rotator with $P = 152.62$~h, which are (2776) Baikal and (8315) Bajin, respectively. All the others, namely (12722) Petrarca,
(13066) 1991 PM13, (23495) 1991 UQ1, (49863) 1999 XK 104, and (70184) 1999 RU3, are rather fast rotators, with periods between 3 and 10~h.

%-------------------------------------------------------------------
\subsection{Spin pole directions}
By combining new and literature data, we successfully determined the shapes and spin states for 55 asteroids that belong to the nominal population of the primitive primordial family of the inner main belt \citep{Delbo2017}. This corresponds to 51\% of the population in the sliver between the left-wing border of the Polana family and the primordial family (see Fig. \ref{fig:ObsByUs}). In particular, we calculated 25 new complete asteroid models, 20 revised and 4 new partial models (see Tables~\ref{tab:NewModels}~and~\ref{tab:RevisedModels}).
Specifically, 34 asteroids have retrograde rotation, while 21 are prograde. The case of (2575) Bulgaria  is presented as an example of model fits to the dense and sparse photometric data in Figs.~\ref{fig:2575} and \ref{fig:2575sp}, respectively. Figure~\ref{fig:shapes} shows an example of four shape models derived from our analysis.

%-------------------------------------------------------------------
\subsection{Identification of interlopers}
In order to proceed and study the spin states of the inner main belt primordial family, we need to eliminate any potential interlopers beforehand. These are defined as objects that may reside inside the V-shape of a family and/or that are grabbed by the HCM methods in the same cluster, but their physical properties, such as the geometric visible albedo $p_\mathrm{V}$ and spectral class, are totally different from the bulk composition of the family \citep{Nesvorny2015}. These asteroids that are clearly non-family members, may belong to the background population or to other nearby families.

In our case seven asteroids have reported spectral (visible or near-infrared) or spectrophotometric (e.g. SDSS, ECAS, and MOVIS) data that do not match the primordial family composition, which is composed primarily of dark albedo and featureless spectra. Specifically, asteroids (1806) Derice, (2171) Kiev, (2575) Bulgaria, and (6125) Singto are classified as S-complex asteroids; (2768) Gorky as an A-type; and (5524) Lecacheux and (15415) Rika   as V-types in any of the main classification schemes of Tholen \citep{Tholen1989}, Bus \citep{Bus2002}, and Bus\&DeMeo \citep{DeMeo2009}. Of these, asteroids (1806), (2171), (2575), and (2768) had been already assigned to the nearby bright family of Flora that consists of S-complex asteroid members, which leaves no doubts that they are interlopers of the low-albedo primordial family of \citet{Delbo2017}. Although asteroids (2536) Kozyrev and (2705) Wu have no spectral or spectrophotometric information, their $p_\mathrm{V}$ values are moderate and beyond the 12\% that is generally defined as the separation between the S and C spectroscopic complexes \citep{Delbo2017}. In the studied sample there are seven asteroids (220, 428, 917, 1244, 1544, 1700, 3633) whose $p_\mathrm{V}$ is in agreement with the primordial family but are classified as members of the X-complex. Using only the visible part of the spectrum (or spectrophotometric data) is not always sufficient to distinguish the slope between the X- and C-complex asteroids, the latter being the main components of the primordial family. So, although these seven asteroids could   indeed be interlopers, at this stage we cannot definitely exclude them from the family. 

On the other hand, in the population that is studied in this work there are asteroids that had been previously assigned using the clustering methods to the other nearby families of Flora and Vesta \citep{Nesvorny2015}, such as the asteroids (428) Monachia, (4524) Barklajdetolli, (2839) Annette, and (3633) Mira. 
However, their low (<10\%) $p_\mathrm{V}$ values and their featureless spectrophotometric data are in contrast to this assignment, and therefore remain in the primordial family. 

The above analysis indicated nine interlopers in the sample of 55 studied objects. From these 46 confirmed asteroid members of the primordial family, 31 asteroid models (67\%)  have retrograde rotation and 15 prograde,   including   the partial solutions. Excluding the seven X-complex asteroids the abundance of retrograde asteroids reaches 72\%. The distribution of the sense of rotation (i.e. prograde or retrograde) within the family is presented in Fig.~\ref{fig:ObsByUs}~(Panel B). Additionally, Fig.~\ref{fig:spd} illustrates the distribution of spin axis directions. Table~\ref{tab:PhysAst} presents all the objects studied in this work, indicating their prograde or retrograde spin along with their physical properties. Diameters ($D$) and geometric albedos ($p_\mathrm{V}$) were calculated as the weighted averages of all the available measurements in the literature and were retrieved from the Minor Planet Physical Properties Catalogue. All uncertainty-weighted averages use 1/$\sigma^2$ as weights, where $\sigma$ is the error of each measurement. 

%-------------------------------------------------------------------

\section{Discussion} \label{sec:discussion}

\subsection{The cases of (2171) Kiev, (7132) Casuli and (2705) Wu}

Most of the asteroids studied in our sample are found to be single objects with no evidence of any close companion body or satellite. However, two asteroids in our sample host a satellite, while another is probably  a non-principal-axis (NPA) rotator. 

The first two cases are (2171) Kiev and (7132) Casuli, which have been reported as binaries by \cite{loera2020lightcurve} and \cite{franco2020asteroid}, respectively. Our dense photometric data confirm that (2171) Kiev and (7132) Casuli are indeed binary asteroids; however, the observed light curves are not enough to   further constrain  the orbits of  secondaries. In each case the eclipsing part of the light curve was removed and the CI method was applied to the rotational light curves of the primary body. Thus, the results presented in Tables~\ref{tab:PhysAst} and \ref{tab:NewModels} are only for the primary bodies. Although the derived parameters do not deviate from the rest of the sample, care should be taken given  that the companion body can alter the spin axis and rotational characteristics of the primary body. 

The third case, (2705) Wu, is a slow retrograde rotator, as our analysis and that  of \cite{Durech2020}  have shown by using sparse photometric data. Previous dense-in-time observations have shown that it is possibly a NPA rotator \citep[i.e. tumbler,][]{oey2010wu}. As  has been noted, some deviations from the single periodicity are clearly seen, but not at a conclusive level, while more photometric data are needed to resolve the second period. In this study no further dense-in-time observations were obtained. The characteristic timescale of damping of the excited NPA rotation can be estimated as $\tau_d= P_{[h]}^3/(C^3 \cdot D_{[km]}^2)$~[Gyr] by \cite{harris1994tumbling}, where $C=17 \pm 2.5$~Gyr~$\cdot$~km$^2$/h$^3$. For this asteroid, the damping timescale is estimated to be $12 \pm 5$~Gyr, which is greater than the age of our Solar System. This is   statistically common for NPAs, which have a diameter larger than $\sim 0.4$~km \citep{pravec2005tumbling}. So, if (2705) Wu is a tumbler, the spin solution of our study  related to the sense of rotation can be trusted.

%-------------------------------------------------------------------
\subsection{Distribution of the spin poles}
All asteroid models with retrograde solutions, except the solutions for (933) Susi and (49863) 1999 XK104, have large ecliptic pole latitude values $|\beta| \geq 30^\circ$ with a large predominance towards the YORP end state values approaching $\beta\sim-90^\circ$. The latitude distribution for prograde rotators differs slightly from the retrograde rotators by having more values with $|\beta| \leq 60^\circ$. This is likely due to various resonances acting only on prograde rotators. Similar behavior is also observed in other asteroid families \citep{Hanus2013c}. 

%prograde lambda is strange, comments.
The distribution of ecliptic pole longitudes is bi-modal for prograde rotators (with two peaks, at $\sim 45^\circ$ and $\sim 225^\circ$) and irregular for retrograde asteroids (with two peaks, at $\sim 75^\circ$ and $\sim 255^\circ$). Previous studies have estimated that the longitude distribution for main belt asteroids is uniform with no statistically significant features \citep{kryszczynska2007new,Hanus2011}. On the contrary, more recent studies present an anisotropic longitude distribution with two symmetrical maxima around $\sim 50^\circ$ and $\sim 230^\circ$ and minima around $\sim 140^\circ$ and $\sim 320^\circ$ \citep{Bowell2014a,cibulkova2016distribution}. 

Retrograde and prograde asteroids also have  different period distributions. As Fig.~\ref{fig:Phist} shows, the majority of slow rotators are retrograde. Moreover, the periods of retrograde asteroids have a non-Maxwellian distribution with excesses  at the fast and the slow rotations. This bimodal distribution is in agreement with that of main belt asteroids $<$40~km, as a result of the YORP effect \citep{pravec2000fast,pravec2002asteroid}. The period distribution of prograde asteroids is    irregular,  with small peaks for slow, moderate, and fast rotations. Thus, prograde and retrograde seem to have different spin rate distributions. A simple model by \cite{pravec2008spin} estimated that a uniform distribution for $<$40~km main belt asteroids could happen on large timescales. Moreover, the different spin rate distributions could signify a different YORP evolution \citep[and references therein]{pravec2000fast,pravec2002asteroid, pravec2008spin}.

%-------------------------------------------------------------------

\subsection{Statistical predominance of the retrograde spin poles}

It is possible to test whether the observed predominance of the retrograde spin poles could be due to chance,  created by   random sampling of  an equal-probability population of prograde and retrograde asteroids. In particular, we test the probability of obtaining 15 or fewer prograde rotators from 46 observed asteroids, drawing from a population having an equal probability (0.5) of  being retrograde or prograde:  \begin{equation}
p(\leq\!15,46) = \sum_{j=1}^{15} \binom{46}{j} ~0.5^{j} (1-0.5)^{46-j}.
\label{E:pro}
\end{equation}
An evaluation of Eq.~\ref{E:pro} gives $p(\leq\!15,46)=$ 1.29\%. This shows that our observations rule out the null hypothesis (i.e. there is no statistical predominance of the retrograde spin poles)  at 98.71\% probability.

%-------------------------------------------------------------------
\subsection{The YORP reorientation timescale for the primordial family members}

Although a predominance of retrograde spinning asteroids is expected in the inward wing of the V-shape of a collisional family, it is also likely to find prograde rotators. During the family lifespan of a few billion years, there are a number of processes that can re-orient the spin vector on an asteroid, such as non-catastrophic collisions, activity, or topographic changes, for example due to mass wasting and/or movement \citep[see e.g.][for a discussion]{paolicchi2016footprints, bottke2015search}. Torques on asteroids due to the unbalanced emitted and reflected radiation \citep[which cause the YORP effect][]{Rubincam2000} are very sensitive to topographic features, and so are the strength and sign of the YORP effect \citep[which  can be reversed;][]{statler2009extreme}. \cite{statler2009extreme} theoretically showed that YORP is sensitive to small surface features, and  hence even a small impact or mass movement could alter the shape sufficiently and change the sense of pole--period evolution. Moreover, as the YORP effect drives some of the family members towards the critical threshold for fast rotation, topographic instability might easily occur, leading to new YORP coefficients, which can drive the spin-pole in the opposite direction. The older a family is, the higher  the cumulative probability is that some family members may had undergone shape changes or received non-catastrophic impacts, hence the higher the probability of detecting prograde rotators in the wing of the V-shape with a predominance of retrograde rotators.

Given the above, we used the model of \citet{vokrouhlicky2006yarkovsky} to estimate the evolution of the spin vectors of the family member asteroids. The model of \citet{vokrouhlicky2006yarkovsky} takes into account the change in orbital semi-major axis due to the Yarkovsky effect and the changes in the rotation rate $\frac{d\omega}{dt}$ and spin axis obliquity\footnote{Obliquity ($\epsilon$) is defined as the angle between the equatorial and orbital planes of an asteroid. For small orbital inclinations, an obliquity of $0^\circ$ is   $\sim 90^\circ$, and $180^\circ$ is $\sim -90^\circ$.}
$\omega \frac{d\epsilon}{dt}$ of the spin axis due to the YORP torques. In particular, the values of $\frac{d\omega}{dt}$ and $\omega \frac{d\epsilon}{dt}$ are multiplied by a constant named $C_{YORP}$, which \cite{bottke2015search} have proposed to be between 0.5 and 0.7. However, \citet{vokrouhlicky2006yarkovsky} found that values of $C_{YORP} \sim 1$ could produce model asteroid families that well represent the real ones. Hence,  for simplicity, we assumed here that $C_{YORP} = 1$. The values of $\frac{d\omega}{dt}$ and $\omega \frac{d\epsilon}{dt}$ are functions of the orbital and physical parameters of asteroids, as described by \citet{vokrouhlicky2006yarkovsky}, which we follow hereafter.

Following \citet{vokrouhlicky2006yarkovsky}, we modelled the evolution of the 46 family members for which we have a pole solution. We assumed they all started with an initial retrograde spin direction, which we randomly assigned uniformly between 90$^{\circ}$ and 180$^{\circ}$ obliquity. We initialised model asteroids with the known diameters and current spin periods. We evolved the model with a time step of 10 Myr for 4 Gyr. At each time step  the values of $a$, $\omega$, and $\epsilon$ were updated by summing their respective time derivatives multiplied by the step in time. At each time step we also evaluated the probability that an asteroid could suffer a non-catastrophic collision capable of changing its spin axis. This probability is given by the 10 Myr time step divided by the re-orientation timescale, $\tau_{reor}$, which is  estimated as

\begin{equation}
    \tau_{reor}=0.845 \left( \frac{5}{P_{[h]}} \right)^{5/6} \left( \frac{D_{[km]}}{2} \right)^{4/3}~[Gyr]
    \label{eq:Treor}
.\end{equation}

Then, for each asteroid (at each time step), we extracted a random number, uniformly distributed between 0 and 1, and we re-oriented the spin state of the asteroid when the random number was smaller than said probability. The spin axis re-orientation is performed by picking a new random direction of the obliquity uniformly distributed between 0$^{\circ}$ and 180$^{\circ}$.

After the model was completed, we counted how many of the initially retrograde asteroids became prograde rotators due to spin evolution. We ran the model 10,000 times and we found that the probability of having 15 or fewer prograde rotators of the initial 46 retrograde objects is 13.5\%; on average, we found 11.4 prograde asteroids.

However, \cite{Delbo2017} showed that D$\geq$35 km asteroids could be primordial objects that accreted as planetesimals from the dust of our protoplanetary disk. It is therefore possible that some of them are within the inward wing of the V-shape of the primordial family. Hence, we also considered   the above model only for D$<$35 asteroids, of which 14 are prograde rotators and 29 retrograde rotators. In this case we found a 20\% probability of  having 14 or fewer prograde rotators of the initial 43 retrograde objects is 20\%.

We can conclude that it is possible to observe the current mix of retrograde and prograde rotators within the inward wing of the 4 Gyr  collisional family.

%-----------------------------------------------------------------

\section{Conclusions}
We carried out a campaign of photometric observations of those asteroids that have been classified as   members of one of the oldest collisional (primordial) families in the Solar System \citep{Delbo2017}. We constructed photometric time series,  the light curves,  for 49 asteroids. This corresponds to 46\% of  the members of the primordial family. We combined our light curves with those from the literature,  and with sparse-in-time photometry in order to create multi-epoch photometric datasets to be used as inputs for the convex inversion method. We obtained 49 new and revised shape models and their spin vector solutions. We combined this with the literature spin vectors (for six objects). 

We reassessed the albedo values for the observed asteroids and studied their literature spectra. This allowed us to find nine interlopers among the initial list of family members of \citet{Delbo2017}. After  removing these interlopers, we find that 31 and 15 out of the remaining 46 asteroids are retrograde and prograde, respectively. We show that this predominance of retrograde compared to prograde asteroids is very unlikely (1.29\% probability) to be due to sampling a distribution of objects with equal probability of being prograde and retrograde. This corroborates the hypothesis that the statistical predominance of the retrograde spin poles is due to a physical process, as  was claimed by \citet{Delbo2017}, namely formation as collisional fragments of a common parent body, a subsequent dynamical evolution driven by the Yarkovsky effect. 

\begin{acknowledgements}
MD and CA acknowledge support from ANR “ORIGINS” (ANR-18-CE31-0014). This work is based on data provided by the Minor Planet Physical Properties Catalogue (MP3C) of the Observatoire de la Côte d’Azur. The research of JH has been supported by the Czech Science Foundation through grant 20-08218S. The work of OP has been supported by INTER-EXCELLENCE grant LTAUSA18093 from the Ministry of Education, Youth, and Sports. Support for T.W.-S.H. was provided by NASA through the NASA Hubble Fellowship grant HST-HF2-51458.001-A awarded by the Space Telescope Science Institute (STScI), which is operated by the Association of Universities for Research in Astronomy, Inc., for NASA, under contract NAS5-26555. 
We thank the Las Cumbres Observatory and their staff for its continuing support of the ASAS-SN project. ASAS-SN is supported by the Gordon and Betty Moore Foundation through grant GBMF5490 to the Ohio State University, and funded in part by the Alfred P. Sloan Foundation grant G-2021-14192 and NSF grant AST-1908570. Development of ASAS-SN has been supported by NSF grant AST-0908816, the Mt. Cuba Astronomical Foundation, the Center for Cosmology and AstroParticle Physics at the Ohio State University, the Chinese Academy of Sciences South America Center for Astronomy (CAS-SACA), the Villum Foundation, and George Skestos.
\end{acknowledgements}

\bibliographystyle{aa} % style aa.bst
\bibliography{References.bib, mybib} % your references Yourfile.bib

\begin{appendix}
\section{Supplementary figures}

\begin{figure*}%[!ht]
\begin{center}
\resizebox{1.0\hsize}{!}{\includegraphics{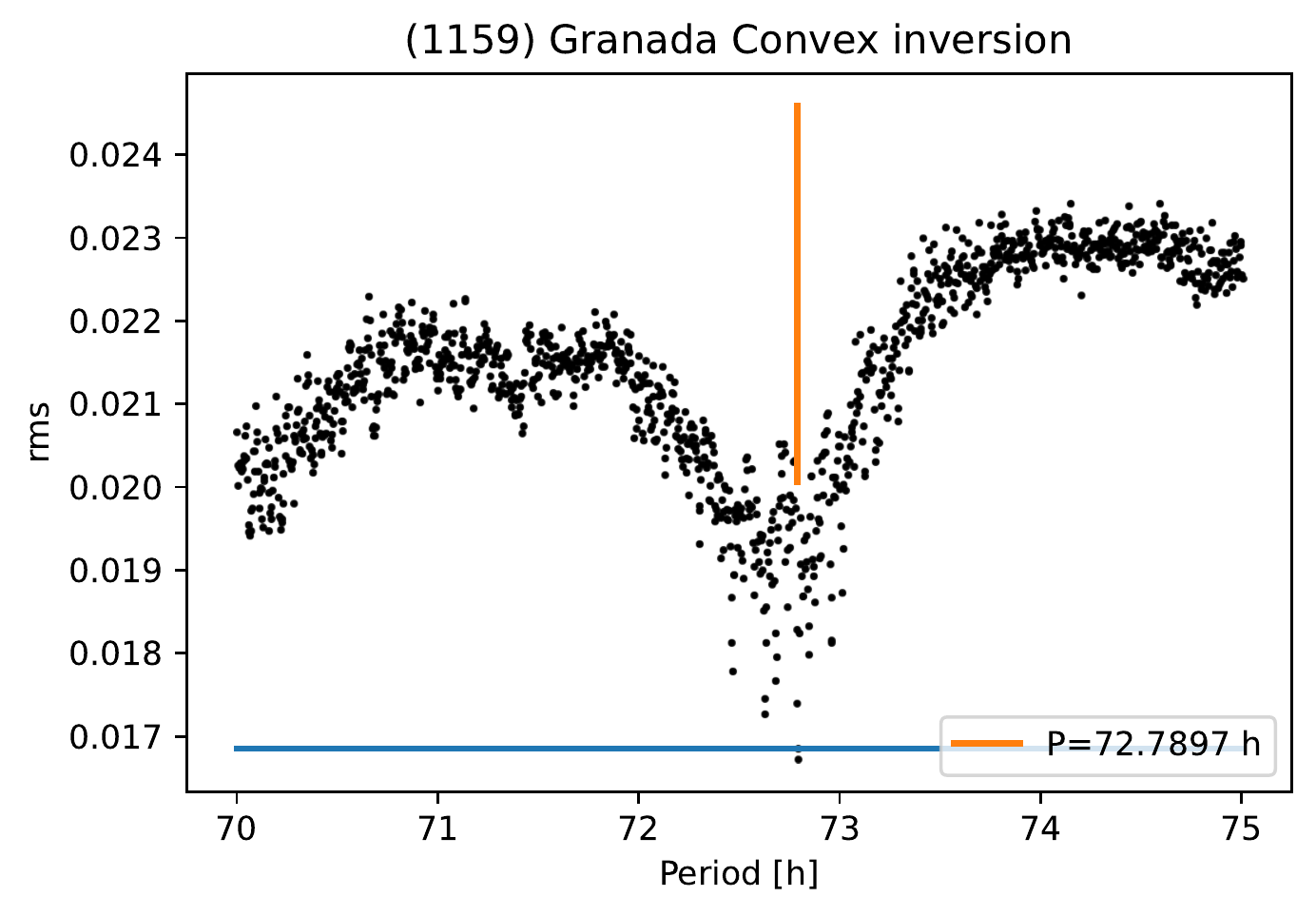}\includegraphics{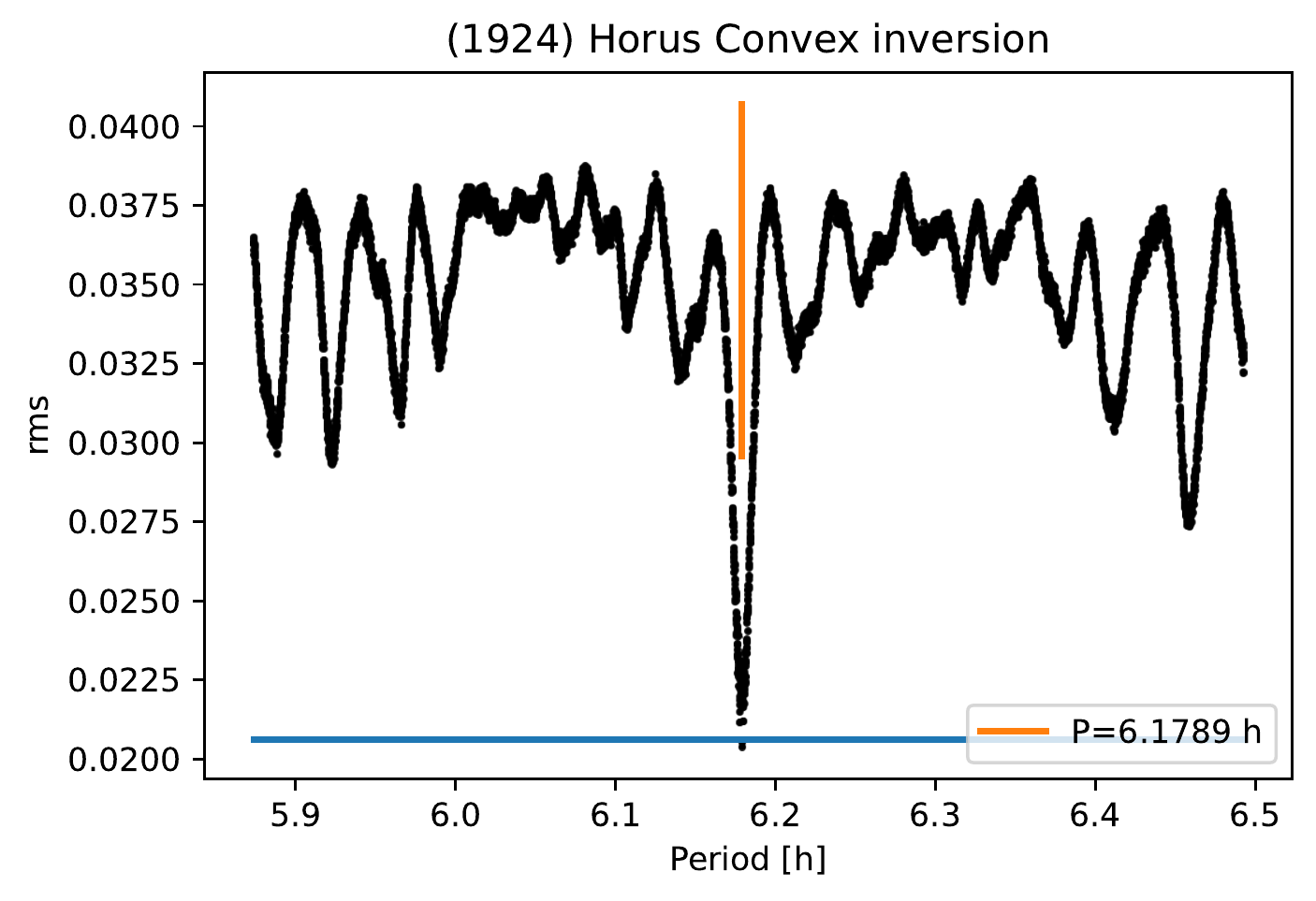}}\\
\resizebox{1.0\hsize}{!}{\includegraphics{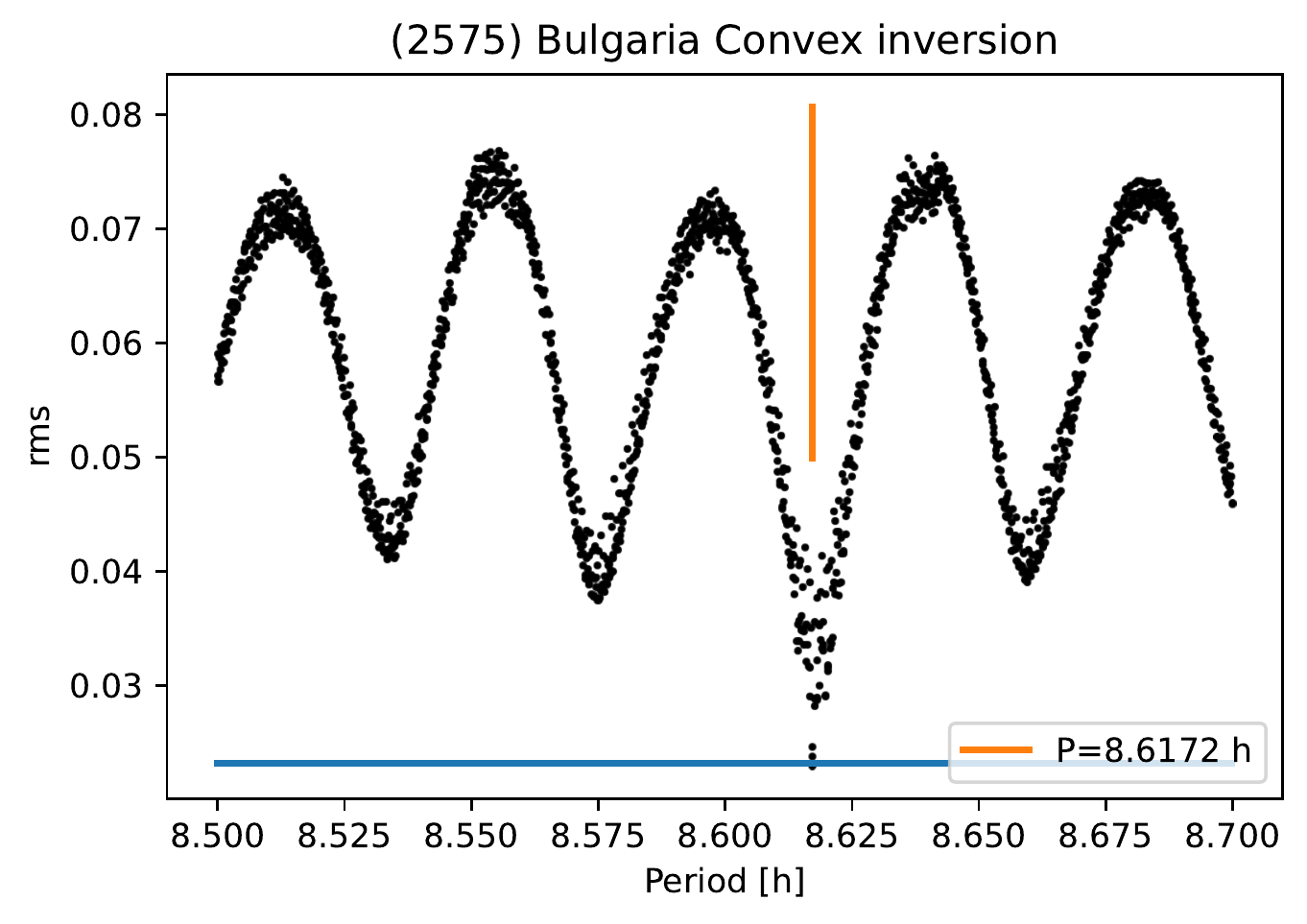}\includegraphics{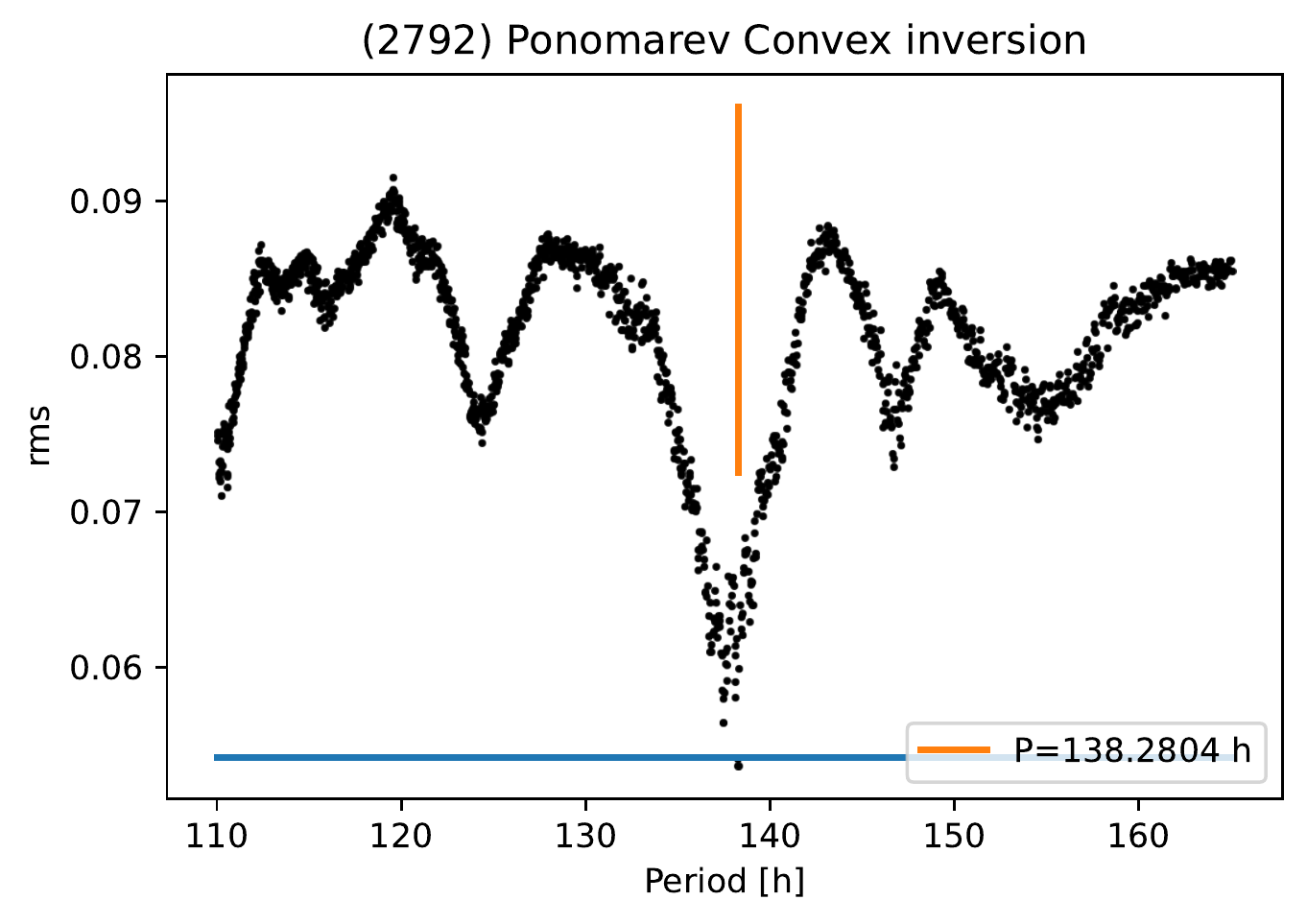}}\\
\resizebox{1.0\hsize}{!}{\includegraphics{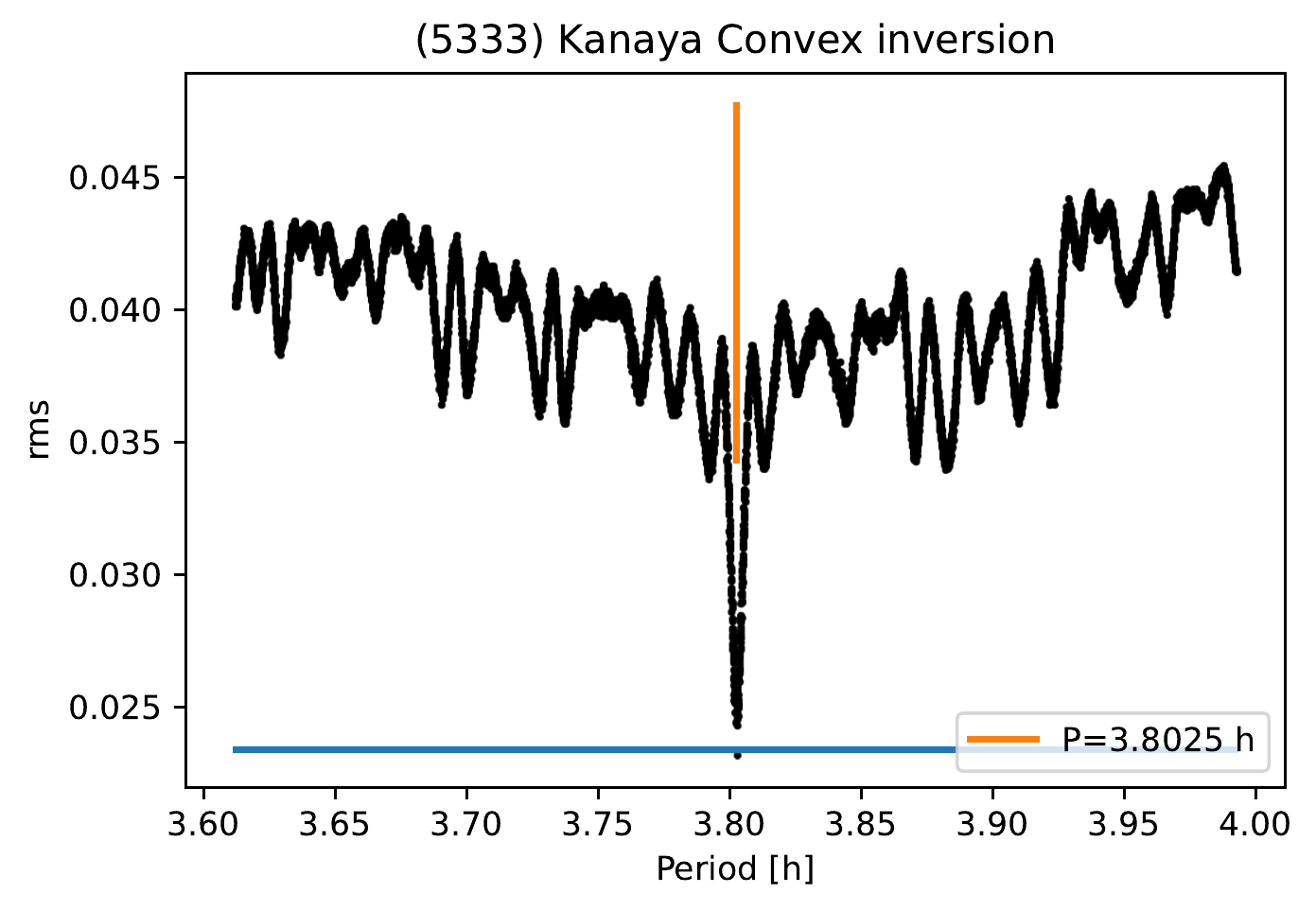}\includegraphics{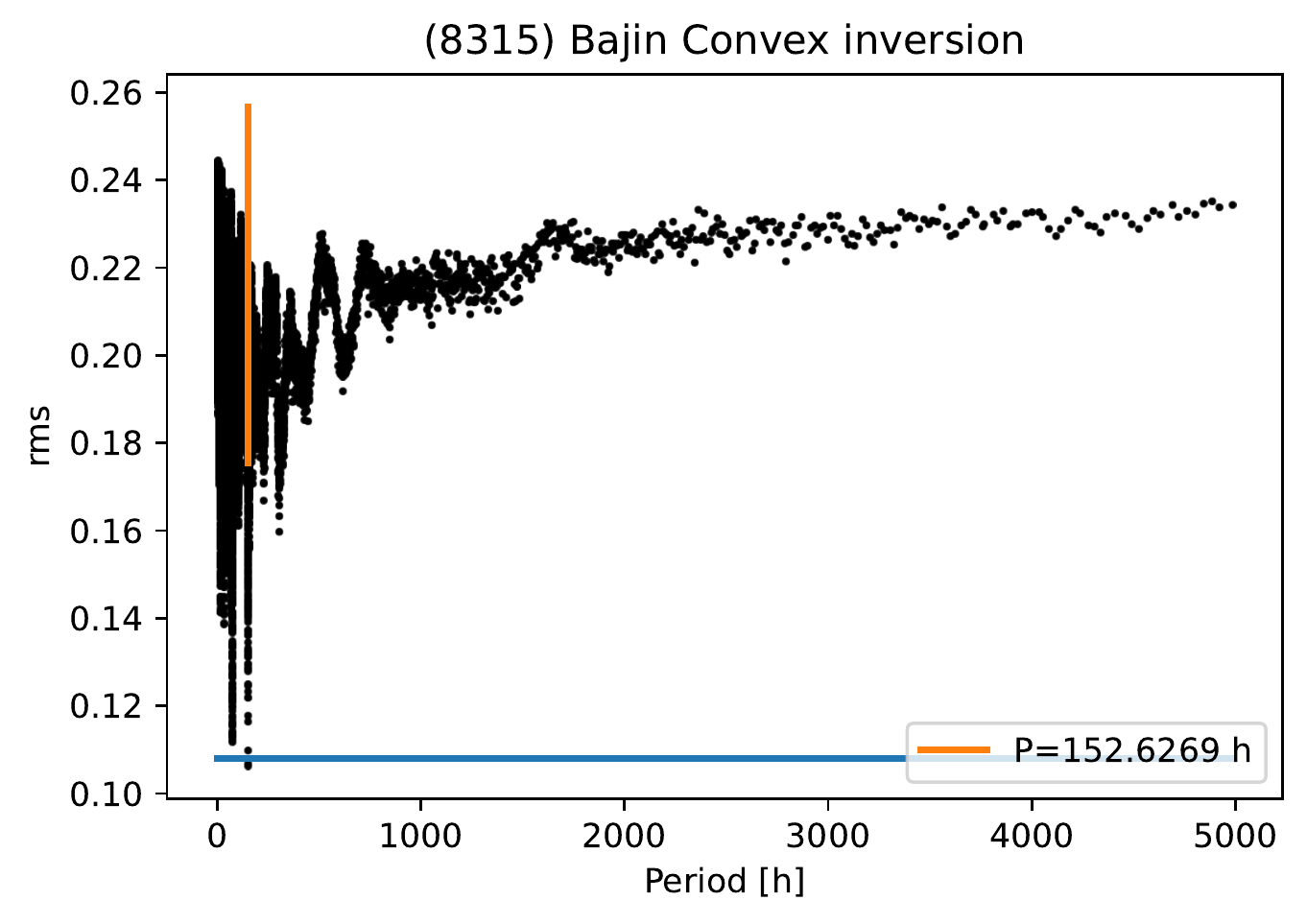}}\\
\end{center}
\caption{\label{fig:periodograms}Typical periodograms of six different asteroids and  their new shape model determinations. Each dot represents one trial run that samples all the local minima at a fixed rotation period (Eq.~\ref{eq:deltaP}) within the searched interval. The vertical lines indicate the best-fit values. The horizontal line represents the $\chi^2$ threshold defined by Eq.~\ref{eq:chi2limit}.}
\end{figure*}

\begin{figure*}%[!ht]
\begin{center}
\resizebox{1.0\hsize}{!}{\includegraphics{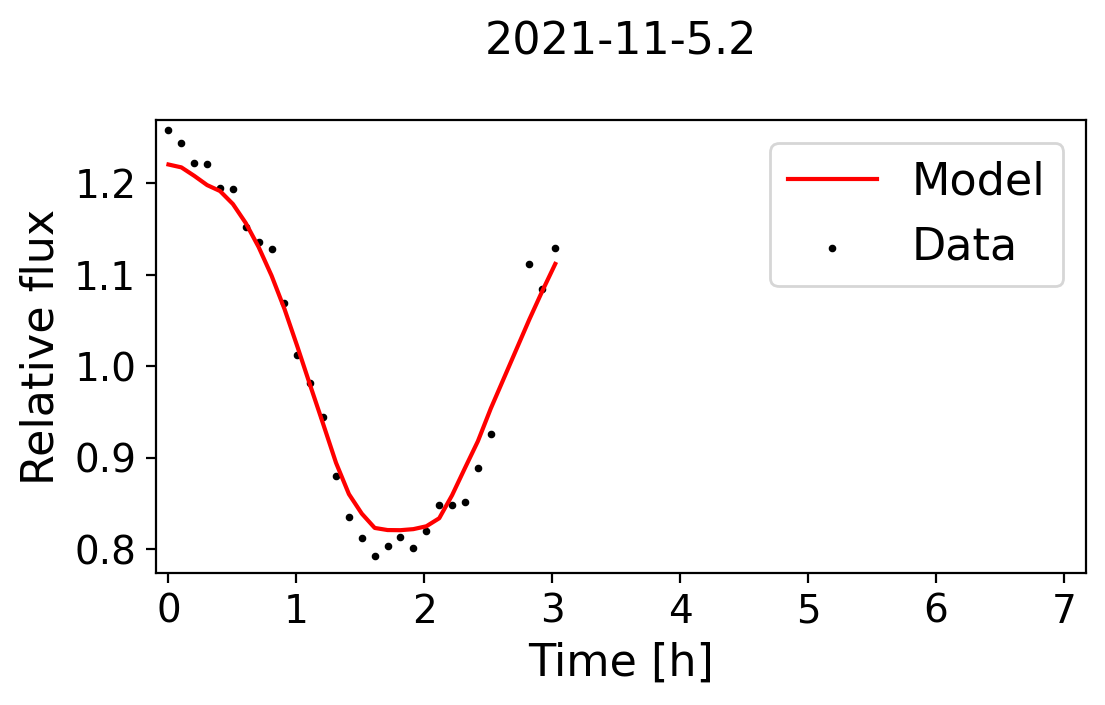}\includegraphics{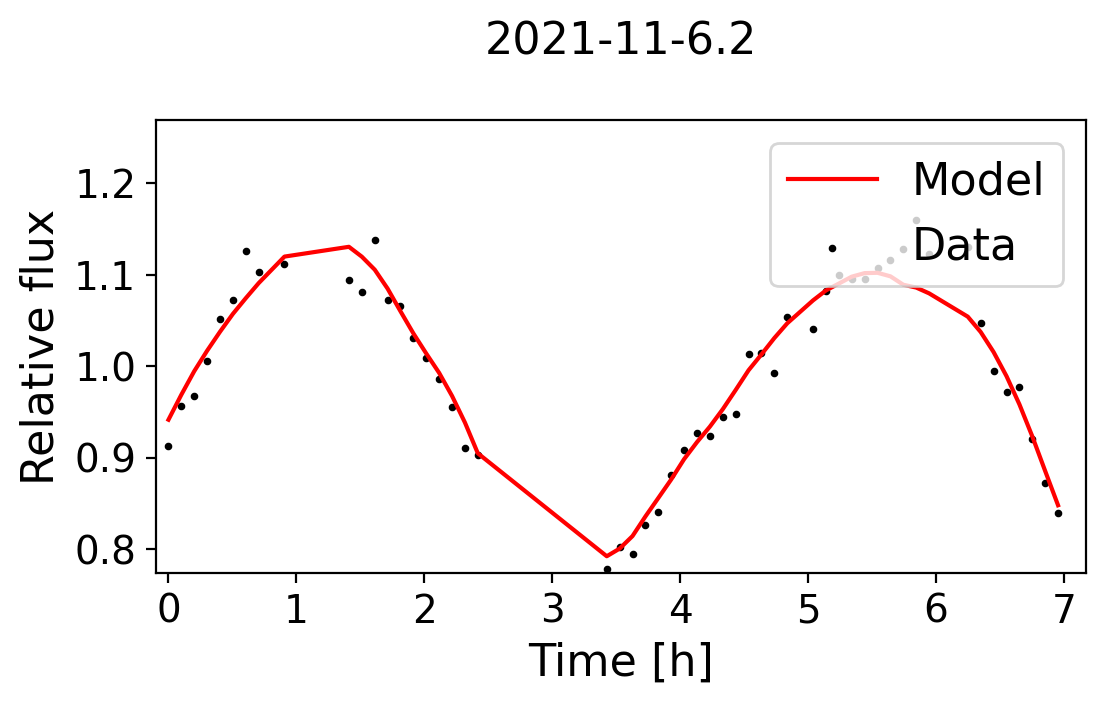}}\\
\resizebox{1.0\hsize}{!}{\includegraphics{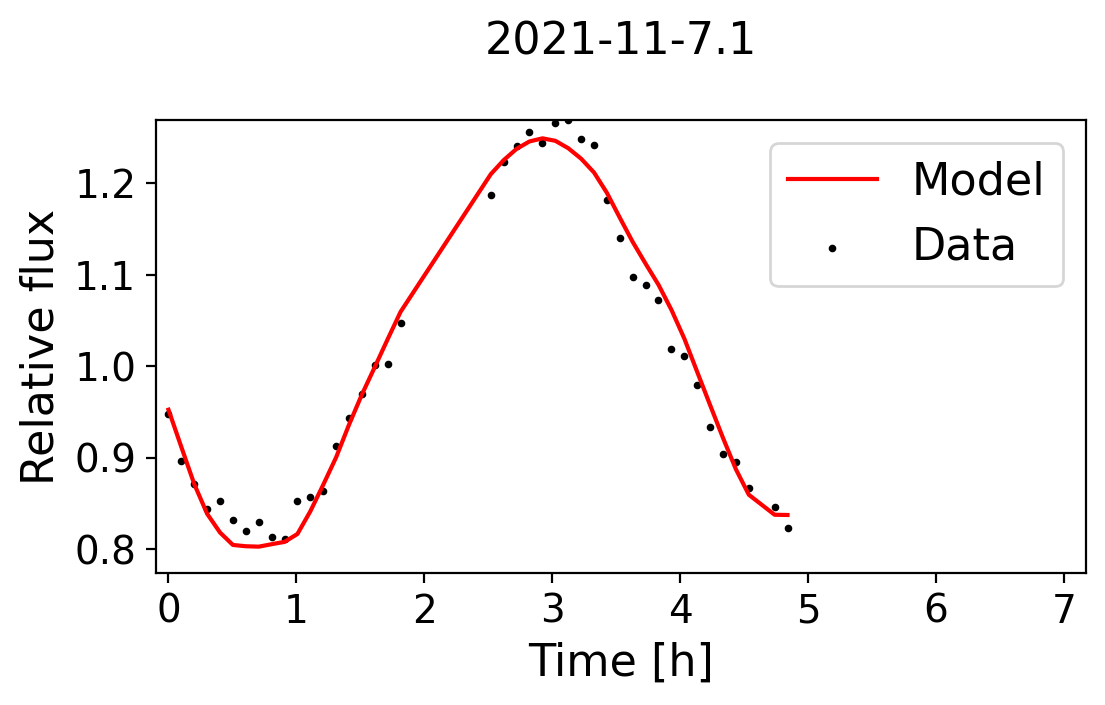}\includegraphics{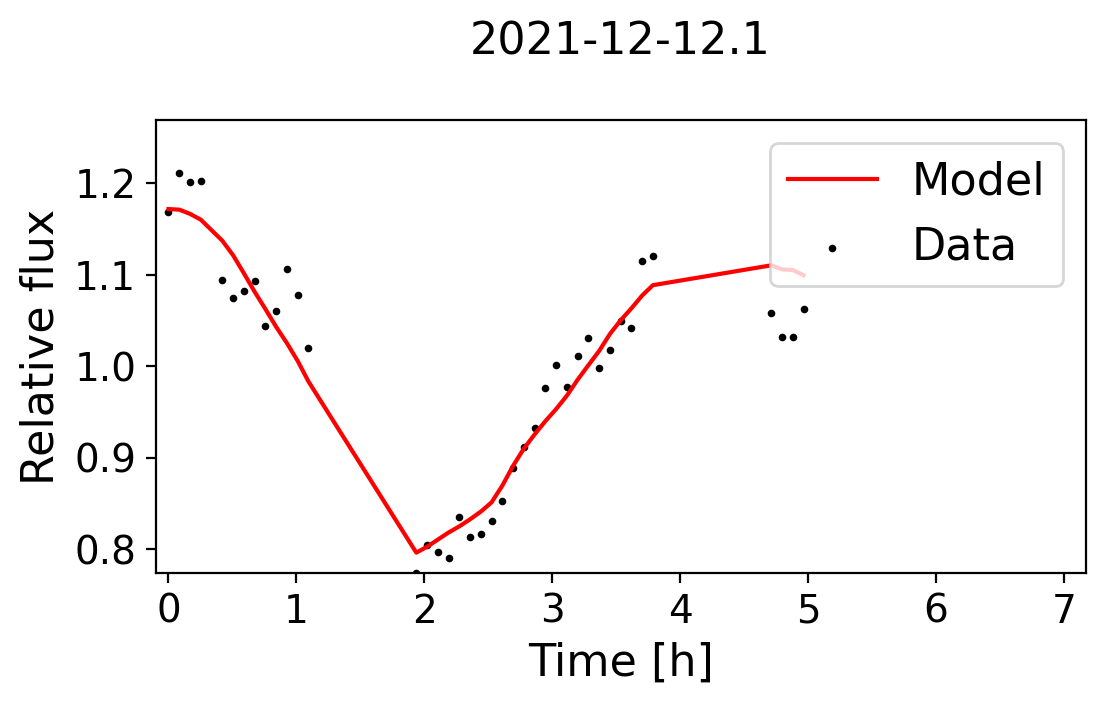}}\\
\resizebox{0.5\hsize}{!}{\includegraphics{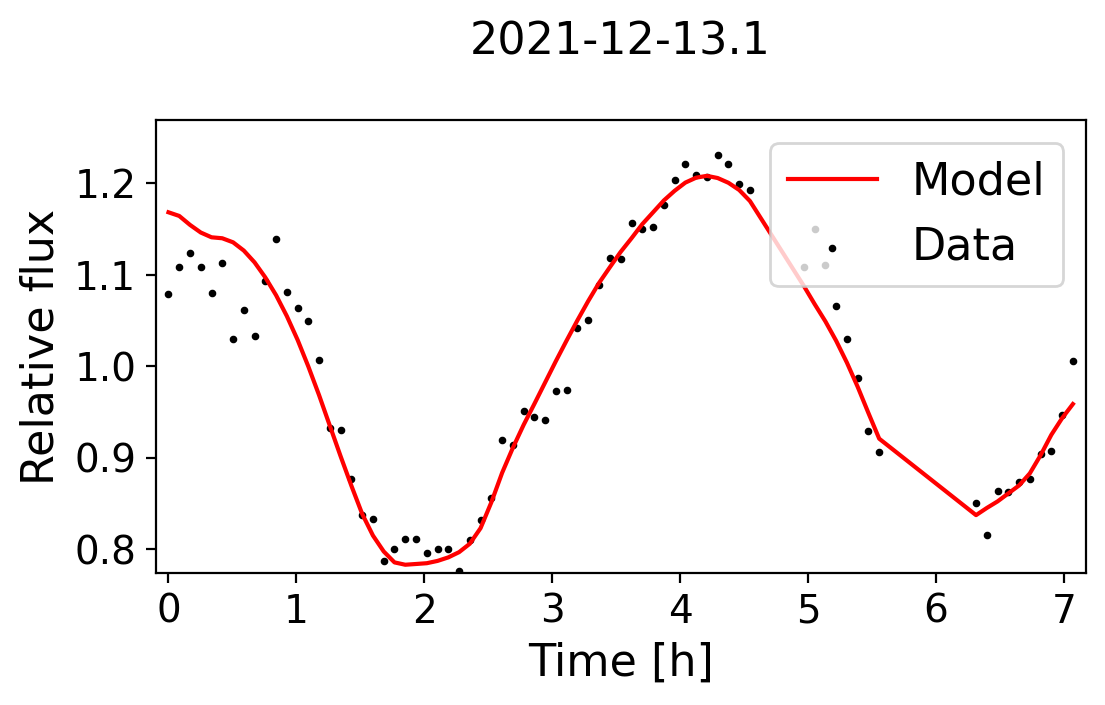}}\\
\end{center}
\caption{\label{fig:2575}Example of fits to the dense light curves for asteroid (2575)~Bulgaria. All five light curves were obtained in the framework of this study (observed by Andrea Ferrero). }
\end{figure*}

\begin{figure*}%[!ht]
\begin{center}
\resizebox{1.0\hsize}{!}{\includegraphics{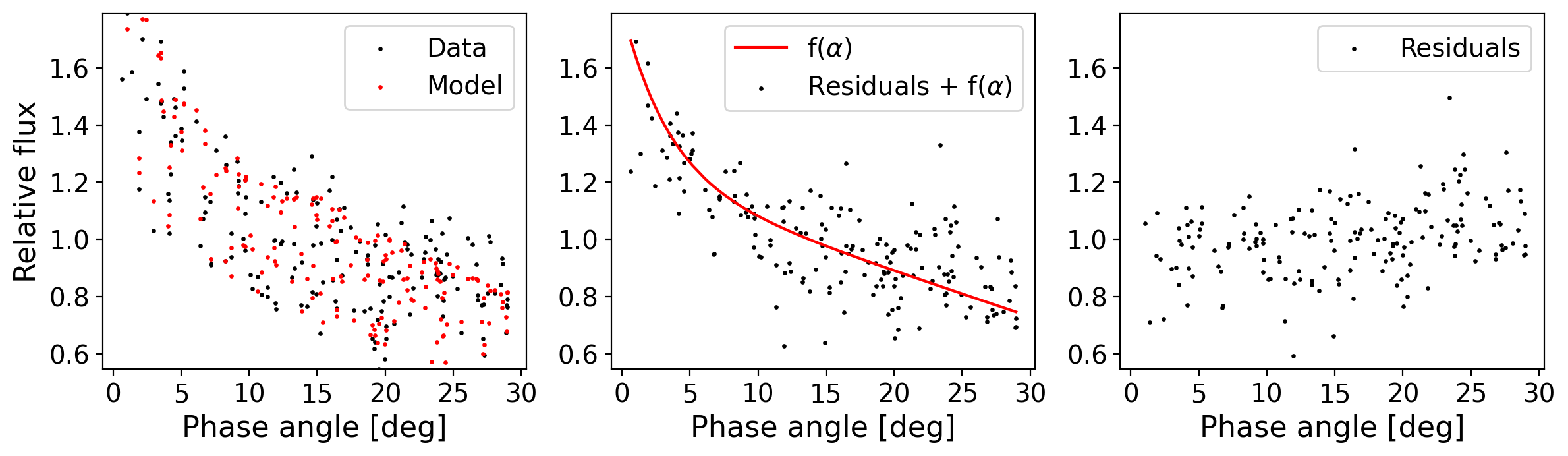}}\\%ASAS V
\resizebox{1.0\hsize}{!}{\includegraphics{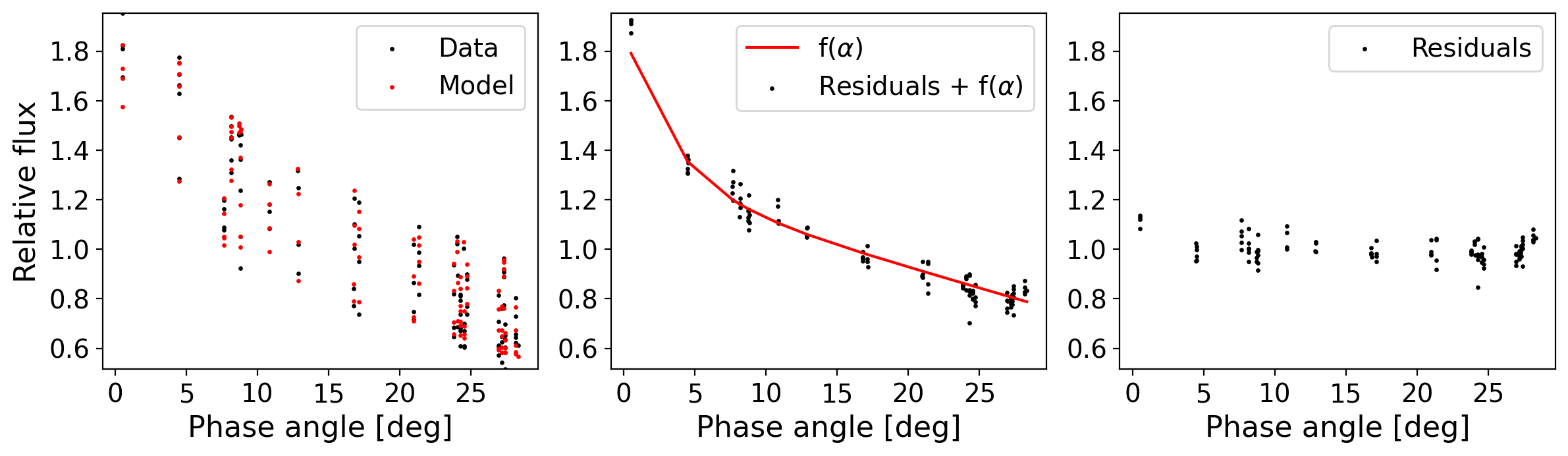}}\\%ASAS g
\resizebox{1.0\hsize}{!}{\includegraphics{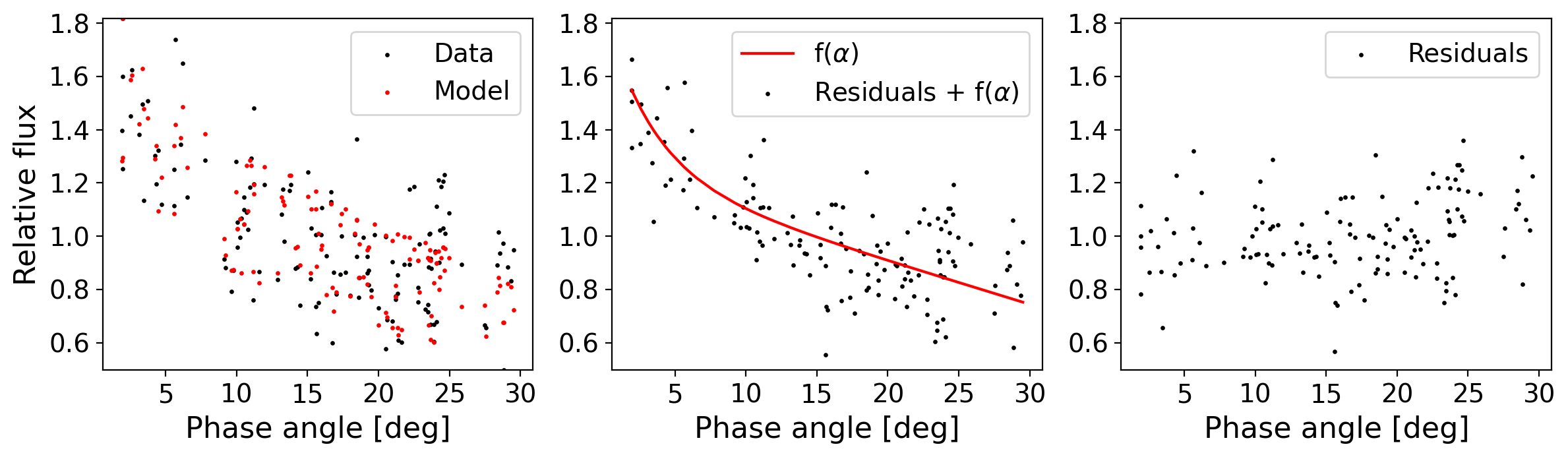}}\\%Atlas c
\resizebox{1.0\hsize}{!}{\includegraphics{./Figures/2575/lcs_fit_sp3.png}}\\%Atlas o
\end{center}
\caption{\label{fig:2575sp}Example of fits to the sparse datasets for asteroid (2575)~Bulgaria. The function f($\alpha$) is our fit of a semi-empirical phase function \citep{kaasalainen2002steroid}. The datasets  (from top to bottom) are ASAS-SN $V$-band, ASAS-SN $g$-band, ATLAS $c$-band, and ATLAS $o$-band.}
\end{figure*}

\begin{figure*}%[!ht]
\begin{center}
\resizebox{0.90\hsize}{!}{\includegraphics{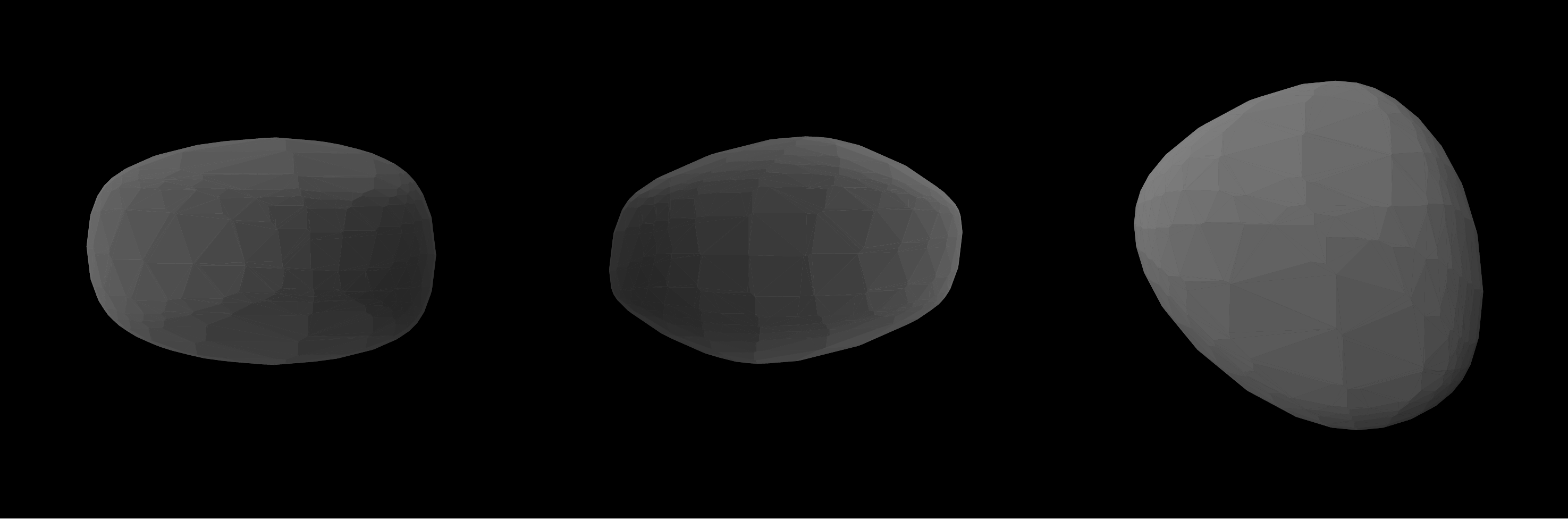}}\\
\resizebox{0.90\hsize}{!}{\includegraphics{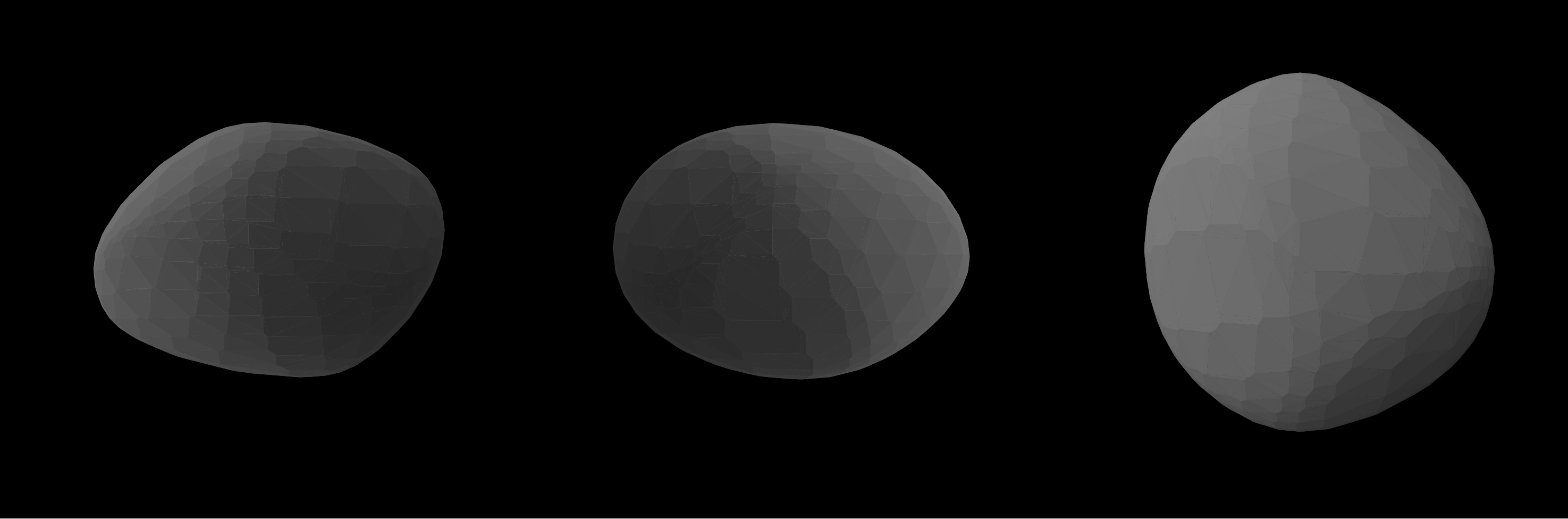}}\\
\resizebox{0.90\hsize}{!}{\includegraphics{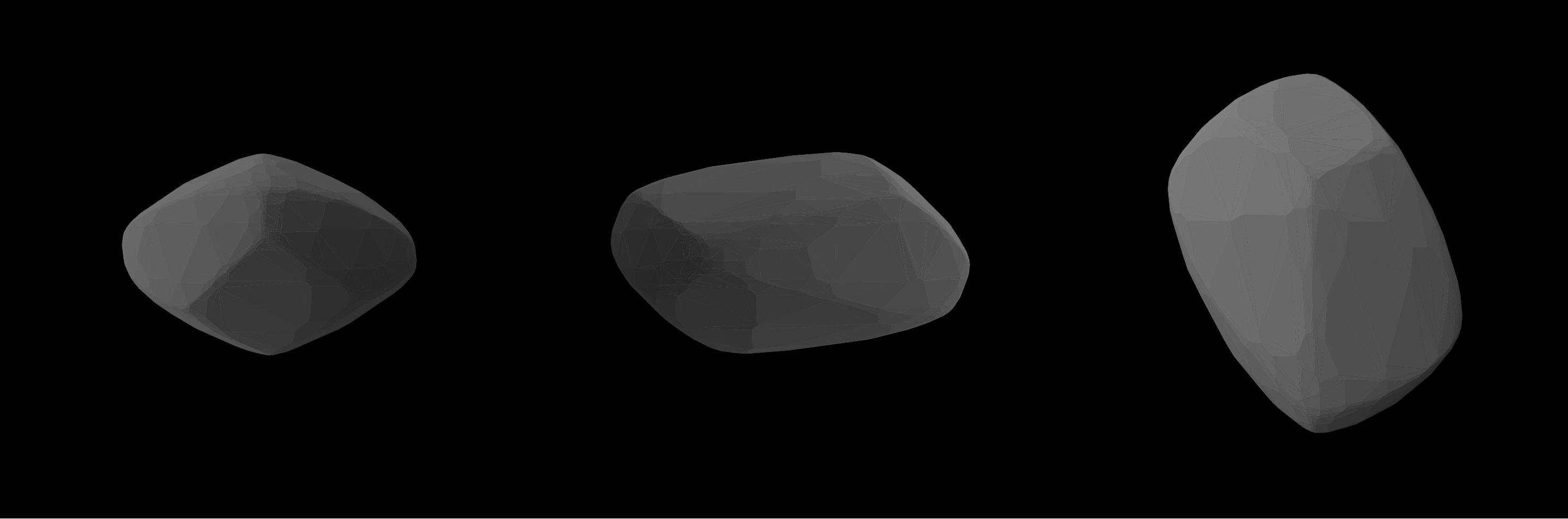}}\\
\resizebox{0.90\hsize}{!}{\includegraphics{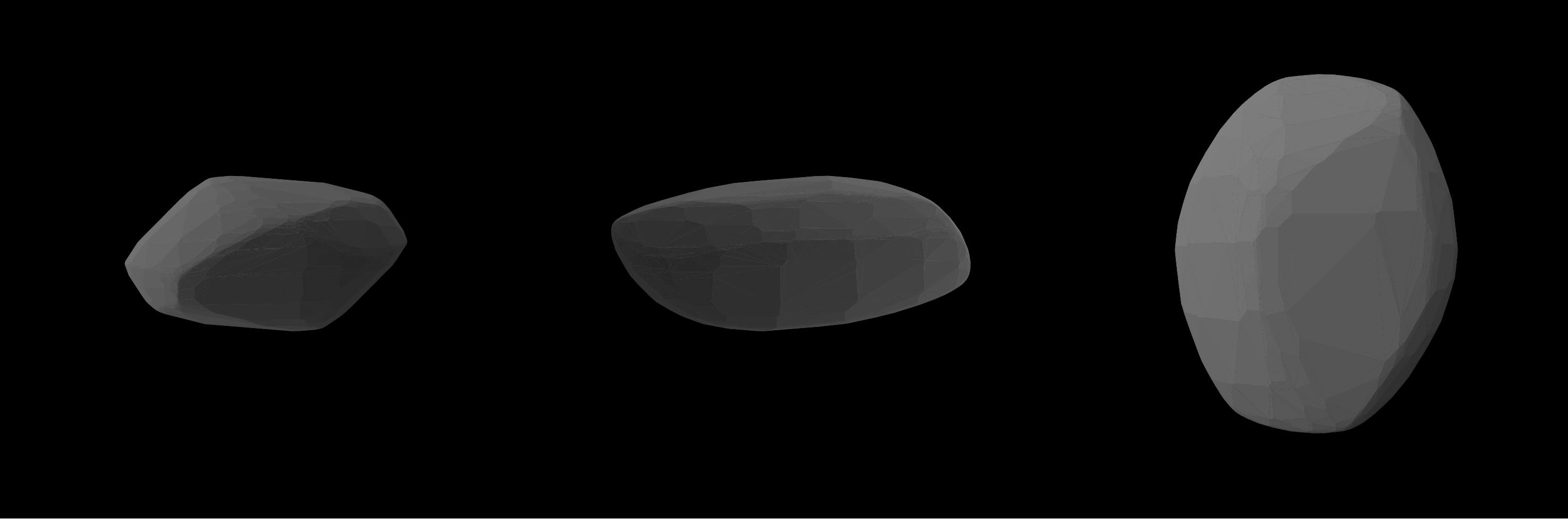}}\\
\end{center}
\caption{\label{fig:shapes}Examples of several shape models. The first panel (top) is the  equatorial view with a 90$^\circ$ rotation offset; the third panel is north pole-on view. The shape models correspond to the first pole solution ($\lambda_1$,$\beta_1$) for each asteroid. The asteroids (from top to bottom) are (1159)~Granada, (1700)~Zvezdara, (2575)~Bulgaria, and (2792)~Ponomarev.}
\end{figure*}

\section{Tables}

\begin{table*}[!ht]
    \caption{Asteroids whose spin poles are known from the literature.}
    \centering
    \begin{tabular}{rlcccccc}
    \hline \hline
    \multicolumn{2}{c}{Asteroid} & $P$ & $\lambda_1$ & $\beta_1$ & $\lambda_2$ & $\beta_2$ & Original model \\
    Number & Name/Designation & (h) & (deg) & (deg) & (deg) & (deg) & publication \\
    \hline
220   & Stephania       & 18.209    & 26    & $-$50         & 223   & $-$62   & \cite{Hanus2013a}\\
249   & Ilse            & 84.995    & 2     & 85            & 222   & 41      & \cite{Hanus2016a}\\
282   & Clorinde        & 49.36     & 353   & $-$66         & 184   & $-$47   & \cite{Durech2020}\\
370   & Modestia        & 22.5411   & $-$   & $-$50 $\pm$ 9 & $-$   & $-$     & \cite{Durech2020}\\
428   & Monachia        & 3.63360   & $-$   & 51 $\pm$ 9    & $-$   & $-$     & \cite{Durech2020}\\
933   & Susi            & 4.6224    & 301   & $-$10         & 125   & $-$15   & \cite{Durech2020}\\
1216  & Askania         & 6.53713   & $-$   & 44 $\pm$ 14   & $-$   & $-$     & \cite{Durech2020}\\
1244  & Deira           & 216.98    & 107   & $-$56         & 314   & $-$46   & \cite{Hanus2016a}\\
1705  & Tapio           & 25.544    & 265   & $-$48         & 106   & $-$57   & \cite{durech2018asteroid}\\
2012  & Guo Shou$-$Jing & 228.33    & $-$   & $-$59 $\pm$ 18 & $-$  & $-$     & \cite{Durech2020}\\
2536  & Kozyrev         & 7.189     & 257   & 16            & 79    & 18      & \cite{Durech2020}\\
2705  & Wu              & 150.8     & 356   & $-$81         & 138   & $-$55   & \cite{Durech2020}\\
2772  & Dugan           & 235.72    & $-$   & $-$58 $\pm$ 20 & $-$  & $-$     & \cite{Durech2020}\\
2839  & Annette         & 10.4609   & 154   & $-$36         & 341   & $-$49   & \cite{Hanus2013a}\\
4231  & Fireman         & 339.5     & 72    & $-$43         & 258   & $-$36   & \cite{durech2019inversion}\\
4524  & Barklajdetolli  & 965.9     & $-$   & 49 $\pm$ 16   & $-$   & $-$     & \cite{Durech2020}\\
5081  & Sanguin         & 10.26460  & $-$   & $-$49 $\pm$ 5 & $-$   & $-$     & \cite{Durech2020}\\
5524  & Lecacheux       & 8.41706   & $-$   & $-$57 $\pm$ 9 & $-$   & $-$     & \cite{Durech2020}\\
5924  & Teruo           & 9.9918    & 340   & $-$44         & 164   & $-$34   & \cite{durech2019inversion}\\
6125  & Singto          & 10.2642   & $-$   & 43 $\pm$ 17   & $-$   & $-$     & \cite{Durech2020}\\
9723  & Binyang         & 12.388    & $-$   & 55 $\pm$ 2    & $-$   & $-$     & \cite{durech2018asteroid}\\
20771 & 2000 QY150      & 8.3014    & 4     & $-$47         & 172   & $-$48   & \cite{durech2019inversion}\\
28736 & 2000 GE133      & 4.6544    & 249   & $-$52         & 134   & $-$84   & \cite{Hanus2016a}\\
30596 & Amdeans         & 23.134    & 114   & 35            & 294   & 37      & \cite{durech2018asteroid}\\
59072 & 1998 VV9        & 7.2982    & 41    & 40            & 223   & 30      & \cite{durech2018asteroid}\\
\hline
    \end{tabular}
    \tablefoot{$P$ is the sidereal period, and $\lambda$ and $\beta$ the ecliptic longitude and latitude of the spin axis, respectively.}
    \label{tab:LitPoles}
\end{table*}

\begin{table*}[!ht]
    \caption{Photometric observations of targets in the frame of  the \textit{Ancient Asteroids} campaign.}
    \label{tab:ObsByUs}
    \centering
    \begin{tabular}{rlcccccc}
    \hline \hline
    \multicolumn{2}{c}{Asteroid} & P$_{synodic}$ & Observing season & N$_{LC}$ & N$_{app}$ & Site\tablefootmark{*} \\
    Number & Name/Designation & (h) & & & &  \\
    \hline
     282 & Clorinde     & 49.352(4)\tablefootmark{a}    & Nov 2020 - Mar 2021   & 7+20      & 1 & BSA+Lowell \\
     370 & Modestia     & 22.5299(1)\tablefootmark{c}   & Aug 2021 - Oct 2021   & 6+1         & 1 & NOAK+OBdB \\
     428 & Monachia     & 3.6343(5)\tablefootmark{b}    & Mar 2018               & 1        & 1 & BE600\\
     853 & Nansenia     & 7.931(2)\tablefootmark{d}     & Dec 2020 - Mar 2021   & 15        & 1 & Lowell \\
     917 & Lyka         & 7.8838(3)\tablefootmark{d}     & Oct 2018 - Nov 2018  & 8        & 1 & OCA(Mont Gros) \\
     933 & Susi         & 4.6222(4)\tablefootmark{f}    & Feb 2018 - Jan 2021   & 3+5+1       & 2 & BSA+Lowell+BE600 \\
    1159 & Granada      & 77.28(5)\tablefootmark{d}     & Oct 2021              & 1+2       & 1 & OCA(C2PU)+UOAO \\
    1700 & Zvezdara     & 9.098(2)\tablefootmark{d}     & Nov 2020 - Jan 2021   & 2+11      & 1 & BSA+Lowell \\
    1806 & Derice       & 3.22443(1)\tablefootmark{g}   & May 2021              & 3         & 1 & BSA \\
    1924 & Horus        & 6.177(14)\tablefootmark{h}    & Sep 2021 - Nov 2021   & 6         & 1 & BO \\
    2012 & Guo Shou-Jing& $-$                           & Oct 2021 - Dec 2021   & 15        & 1 & BO \\
    2171 & Kiev         & 3.1714(2)\tablefootmark{i}    & Jan 2022              & 3         & 1 & BO\\
    2259 & Sofievka     & 63.0918(5)\tablefootmark{b}   & Dec 2020 - Jan 2021   & 9         & 1 & BSA \\
    2322 & Kitt-Peak    & 8.460(6)\tablefootmark{j}     & Nov 2020 - Jan 2021   & 4+9       & 1 & BSA+Lowell \\
    2575 & Bulgaria     & 8.618(7)\tablefootmark{k}     & Nov 2021 - Dec 2021   & 5         & 1 & BO  \\
    2768 & Gorky        & 4.5118(7)\tablefootmark{d}    & Dec 2019 - May 2021   & 4+3+9     & 2 & BSA+BO+UOAO \\
    2772 & Dugan       & 235.0(5)\tablefootmark{e}     & Dec 2007   & 1         & 1 & BMO \\
    2773 & Brooks       & 4.838(1)\tablefootmark{d}     & Dec 2018 - Oct 2021   & 3+4         & 2 & OCA(Mont Gros) + BSA \\
    2776 & Baikal       & $-$     & Jan 2007   & 1         & 1 & BMO \\
    2778 & Tangshan     & 3.468(3)\tablefootmark{l}     & Jan 2018 - Oct 2021   & 5+3+6       & 2 & OCA(Mont Gros)+BSA+Lowell \\
    2839 & Annette      & 10.459(5)\tablefootmark{b}    & Apr 2020              & 5         & 1 & UOAO \\
    3633 & Mira         &  19.17(2)\tablefootmark{h}    & Nov 2021 - Jan 2022   & 12        & 1 & BO \\
    3723 & Voznesenskij & 7.9640(85)\tablefootmark{h}   & Oct 2021              & 2+1       & 1 & OCA(C2PU)+Helmos \\
    4231 & Fireman      & 28.0(2)\tablefootmark{d}      & Dec 2020 - Apr 2021   & 19        & 1 & Lowell \\
    4422 & Jarre        & 7.013(1)\tablefootmark{m}     & Apr 2021 - May 2021   & 5+3+9+3   & 1 & BSA+Lowell+BO+UOAO \\
    5081 & Sanguin      & 10.2619(5)\tablefootmark{b}   & May 2021 - Jun 2021   & 6         & 1 & BSA \\
    5333 & Kanaya       & 3.69(5)\tablefootmark{d}      & Jun 2021 - Jul 2021   & 3+1         & 1 & OCA(C2PU)+ChR \\
    6125 & Singto       & 10.2642(1)\tablefootmark{b}   & Jan 2022              & 3         & 1 & BO \\
    6647 & Josse       & 5.9498(3)\tablefootmark{d}     & Aug 2018              & 6         & 1 & OCA(Mont Gros) \\
    7132 & Casulli      & 3.5238(2)\tablefootmark{n}    & Oct 2021              & 3+1       & 1 & OCA(C2PU)+Helmos \\
    9086 & 1995 SA3     & $-$                           & Oct 2019 - Nov 2019   & 22        & 1 & UOAO \\
    9972 & Minoruoda    & 3.4221(2)\tablefootmark{o}    & Oct 2021              & 1         & 1 & Helmos \\
    10542 & Ruckers    & $-$     & Feb 2022 - Mar 2022   & 4        & 1 & BO \\
   15985 & 1998 WU20    & 18.2252(5)\tablefootmark{b}   & Oct 2021              & 4+1       & 1 & OCA(C2PU)+Helmos \\
   25343 & 1999 RA44    & 590.5(5)\tablefootmark{m}     & Jan 2021 - Apr 2021   & 27        & 1 & BO \\
        \hline
    \end{tabular}
    \tablefoot{
    The literature value of the (synodic) rotational period is given, along with the observing log, where N$_{LC}$ is the number of individual light curves obtained by each corresponding observing site given in the last column, while N$_{app}$ is the number of apparitions. The standard error for each value is expressed in brackets, in units of the  last decimal digit quoted.\\
    \tablefoottext{*}{BSA: Astronomical Observatory BSA, Lowell: Lowell Observatory, BO: Bigmuskie Observatory, OCA: Observatoire de la Côte d'Azur - Calern station (C2PU) and Mont Gros station, UOAO: University of Athens Observatory, Helmos: Helmos Observatory  of National Observatory of Athens, NOAK: NOAK Observatory, BE600: BlueEye 600 Observatory, ChR: Pic de Ch\^ateau-Renard Observatory, OBdB: Observatoire du Bois de Bardon, BMO:Blue Mountains Observatory} \\
    \tablefoottext{a}{\cite{bonamico2021determining}}, \tablefoottext{b}{\cite{Pal2020}}, \tablefoottext{c}{\cite{Stephens2011a}},  \tablefoottext{d}{\url{http://obswww.unige.ch/~behrend/page_cou.html}}, \tablefoottext{e}{\url{https://www.asu.cas.cz/~ppravec/}}, \tablefoottext{f}{\url{https://web.archive.org/web/20081004205615/http://www.david-higgins.com/}},
    \tablefoottext{g}{\cite{Stephens2020c}}, \tablefoottext{h}{\cite{Waszczak2015}}, \tablefoottext{i}{\cite{mahlke2021asteroid}}, 
    \tablefoottext{j}{\cite{polakis2021period}}, \tablefoottext{k}{\cite{erasmus2020investigating}}, \tablefoottext{l}{\cite{stephens2019main}}, \tablefoottext{m}{\cite{ferrero2021lightcurves}}, \tablefoottext{n}{\cite{franco2020collaborative}}, \tablefoottext{o}{\cite{cooney2017minoruoda}}      
    }
\end{table*}

%%%%%%%%%%%%%%%%%%%%%%%%%%%  Table A.  %%%%%%%%%%%%%%%%%%%%%%%%%%
\onecolumn
\begin{landscape}
\begin{table*}[!ht]
    \caption{Rotation state properties and summary of the optical dataset for asteroids for which we derived a new shape solution or estimated the sense of rotation.}
    \label{tab:NewModels}
    \centering
    \begin{tabular}{rl ccccc cc ccccccc}
    \hline \hline
    \multicolumn{2}{c}{Asteroid}& $\lambda_1$ & $\beta_1$ & $\lambda_2$ & $\beta_2$ & $P$  & N$_\mathrm{LC}$ & N$_\mathrm{app}$ & N$_\mathrm{689}$ & N$_\mathrm{703}$ & N$_\mathrm{GAI}$ & N$_\mathrm{ASA}$ & N$_\mathrm{ATL}$ & N$_\mathrm{I41}$ & N$_\mathrm{PTF}$\\
    Number & Name/Designation  & (deg) & (deg) & (deg) & (deg) & (h) & & & & & & & & &\\
    \hline
    \multicolumn{16}{c}{New asteroid models}\\
917    &  Lyka            &  44   &  49   &  230   &  37    &  7.88167   &  13  &  1   &  116  &  205  &  9   &  599  &  582  &  26    &  0\\
1159   &  Granada         &  70   &$-$54  &  142   &$-$85   &  72.789    &  5   &  2   &  130  &  196  &  14  &  382  &  508  &  77    &  49\\
1544   &  Vinterhansenia  &  55   &  22   &  238   &  23    &  13.78270  &  0   &  0   &  104  &  215  &  9   &  389  &  341  &  0     &  0\\
1700   &  Zvezdara        &  66   &  62   &  248   &  48    &  9.11217   &  43  &  2   &  116  &  176  &  20  &  354  &  324  &  51    &  0\\
1806   &  Derice          &  42   &  43   &  216   &  38    &  3.223531  &  17  &  2   &  59   &  286  &  0   &  406  &  488  &  20    &  0\\     
1924   &  Horus           &  131  &$-$59  &  315   &$-$54   &  6.17893   &  6   &  1   &  0    &  178  &  0   &  163  &  459  &  29    &  21\\
2171   &  Kiev            &  145  &  54   &        &        &  3.171587  &  7   &  2   &  0    &  277  &  12  &  339  &  451  &  0     &  0\\      
2322   &  KittPeak        &  195  &$-$78  &        &        &  8.46780   &  13  &  1   &  0    &  191  &  0   &  319  &  425  &  0     &  0\\
2575   &  Bulgaria        &  66   &  91   &  246   &  70    &  8.61720   &  5   &  1   &  20   &  205  &  0   &  309  &  445  &  0     &  28\\
2768   &  Gorky           &  54   &  34   &        &        &  4.50814   &  18  &  2   &  37   &  217  &  11  &  151  &  545  &  0     &  0\\
2773   &  Brooks          &  85   &$-$86  &  241   &$-$71   &  4.83726   &  7   &  2   &  0    &  162  &  14  &  110  &  328  &  0     &  0\\
2776   &  Baikal          &  81   &$-$41  &  264   &$-$27   &  3253.5    &  1   &  1   &  0    &  143  &  0   &  312  &  479  &  44    &  0\\
2792   &  Ponomarev       &  72   &$-$58  &  230   &$-$82   &  138.28    &  0   &  0   &  0    &  193  &  0   &  223  &  409  &  92    &  60\\
3633   &  Mira            &  170  &$-$58  &  346   &$-$50   &  19.1832   &  12  &  1   &  0    &  275  &  10  &  124  &  331  &  0     &  29\\
3684   &  Berry           &  96   &$-$57  &  255   &$-$49   &  11.9074   &  0   &  0   &  0    &  186  &  0   &  0    &  332  &  0     &  0\\
3723   &  Voznesenskij    &  75   &$-$35  &  259   &$-$34   &  7.96485   &  3   &  1   &  0    &  185  &  0   &  52   &  292  &  0     &  20\\
5333   &  Kanaya          &  157  &$-$56  &        &        &  3.80247   &  9   &  2   &  0    &  172  &  25  &  210  &  533  &  0     &  0\\
6647   &  Josse           &  38   &$-$68  &  222   &$-$62   &  5.94972   &  6   &  1   &  0    &  165  &  0   &  24   &  271  &  0     &  0\\
8022   &  Scottcrossfield &  186  &  30   &  359   &  40    &  4.00267   &  0   &  0   &  0    &  228  &  18  &  84   &  293  &  0     &  76\\
8315   &  Bajin           &  6    &  71   &        &        &  152.62    &  0   &  0   &  0    &  167  &  0   &  0    &  262  &  74    &  0\\
%9723   &  Binyang         &  36   &  45   &  215   &  38    &  12.38806  &  2   &  1   &  0    &  127  &  0   &  0    &  357  &  0     &  26\\
12722  &  Petrarca        &  143  &  24   &  346   &  41    &  3.538576  &  0   &  0   &  0    &  138  &  0   &  0    &  175  &  41    &  0\\     
15415  &  Rika            &  117  &$-$79  &        &        &  6.36228   &  0   &  0   &  0    &  164  &  0   &  0    &  275  &  0     &  0\\
23495  &  1991 UQ1        &  84   &$-$70  &        &        &  10.50559  &  0   &  0   &  0    &  124  &  0   &  26   &  370  &  0     &  0\\
49863  & 1999 XK104       &  32   &$-$18  &  212   &$-$22   &  8.00795   &  0   &  0   &  0    &  72   &  0   &  0    &  94   &  0     &  0\\
\hline
    \multicolumn{2}{c}{Asteroid} & $\beta_\mathrm{P}$ & $\delta\beta$ &  &  & $P$ & N$_\mathrm{LC}$ & N$_\mathrm{app}$ & N$_\mathrm{689}$ & N$_\mathrm{703}$ & N$_\mathrm{GAI}$ & N$_\mathrm{ASA}$ & N$_\mathrm{ATL}$ & N$_\mathrm{I41}$ & N$_\mathrm{PTF}$\\
    Number & Name/Designation  & (deg) & (deg) &  &  & (h) & & & & & & & & &\\
     \multicolumn{16}{c}{Partial asteroid models}\\
     \hline
2778   &  Tangshan        &$-$65  &  18   &        &        &  3.460693  &  19  &  4   &  0    &  169  &  0   &  130  &  263  &  0     &  0\\      
11975  &  1995 FA1        &  65   &  9    &        &        &  6.21338   &  0   &  0   &  0    &  119  &  0   &  0    &  235  &  0     &  0\\
13066  &  1991 PM13       &$-$57  &  19   &        &        &  3.93765   &  0   &  0   &  0    &  160  &  0   &  90   &  259  &  33    &  0\\
70184  &  1999 RU3        &$-$59  &  14   &        &        &  5.27716   &  0   &  0   &  0    &  150  &  0   &  46   &  320  &  0     &  0\\
\hline
    \end{tabular}
\tablefoot{
The new determinations are listed first, then the cases where we estimated only the ecliptic latitude of the spin axis (i.e. partial models). Listed are the  physical properties: ecliptic longitudes and latitudes of the spin axis directions $\lambda$ and $\beta$ for one or two possible solutions;  the sidereal rotation period $P$;  the mean value of the ecliptic latitude $\beta_\mathrm{P}$  and  1/2 of the range in latitude within the multiple pole solutions $\delta\beta$ (for partial models). The uncertainty of the pole direction is usually about 10\degr\ and of the period is  of the order of the last decimal digit. Also given is information about the light curve dataset: number of dense light curves N$_\mathrm{LC}$ with the number of covered apparitions N$_\mathrm{app}$;   the number of measurements in each sparse dataset (N$_\mathrm{689}$: UNSO-Flagstaff; N$_\mathrm{703}$: CSS; N$_\mathrm{GAI}$: Gaia DR2; N$_\mathrm{ASA}$: ASAS-SN; N$_\mathrm{ATL}$: ATLAS; N$_\mathrm{I41}$: ZTF; N$_\mathrm{PTF}$: PTF).}
\end{table*}
\end{landscape}

%%%%%%%%%%%%%%%%%%%%%%%%%%%  Table A.  %%%%%%%%%%%%%%%%%%%%%%%%%%
\onecolumn
\begin{landscape}
\begin{table*}[!ht]
    \caption{Updated shape models.}
    \label{tab:RevisedModels}
    \centering
    \begin{tabular}{rl ccccc cc ccccccc}
    \hline \hline
    \multicolumn{2}{c}{Asteroid} & $\lambda_1$ & $\beta_1$ & $\lambda_2$ & $\beta_2$ & $P$ & N$_\mathrm{LC}$ & N$_\mathrm{app}$ & N$_\mathrm{689}$ & N$_\mathrm{703}$ & N$_\mathrm{GAI}$ & N$_\mathrm{ASA}$ & N$_\mathrm{ATL}$ & N$_\mathrm{I41}$ & N$_\mathrm{PTF}$\\
    Number & Name/Designation & (deg) & (deg) & (deg) & (deg) & (h) & & & & & & & & &\\
    \hline
220    &  Stephania       &  36   &$-$51  &        &        &  18.2087   &  9   &  2   &  137  &  242  &  12  &  376  &  599  &  30    &  0\\
249    &  Ilse            &  85   &  79   &  266   &  69    &  84.996    &  29  &  3   &  135  &  254  &  9   &  499  &  529  &  0     &  0\\
282    &  Clorinde        &  189  &$-$36  &  346   &$-$59   &  49.3597   &  29  &  2   &  175  &  197  &  13  &  659  &  548  &  0     &  0\\
370    &  Modestia        &  268  &$-$92  &        &        &  22.5403   &  21  &  3   &  176  &  202  &  11  &  564  &  472  &  0     &  0\\
428    &  Monachia        &  98   &  90   &  283   &  53    &  3.633613  &  28  &  16  &  137  &  260  &  25  &  519  &  396  &  0     &  0\\     
933    &  Susi            &  300  &$-$9   &        &        &  4.62241   &  18  &  3   &  114  &  149  &  0   &  385  &  439  &  33    &  0\\
1216   &  Askania         &  62   &  55   &  260   &  66    &  6.53708   &  0   &  0   &  55   &  138  &  0   &  191  &  370  &  0     &  0\\
2012   &  Guo Shou-Jing   &  46   &$-$79  &  230   &$-$76   &  228.29    &  15  &  1   &  0    &  200  &  12  &  83   &  422  &  0     &  0\\
2536   &  Kozyrev         &  80   &  22   &  256   &  21    &  7.18887   &  3   &  1   &  27   &  138  &  14  &  348  &  494  &  0     &  0\\
2705   &  Wu              &  86   &$-$70  &  265   &$-$82   &  150.78    &  0   &  0   &  0    &  174  &  0   &  191  &  429  &  55    &  0\\
2772   &  Dugan           &  157  &$-$71  &        &        &  235.76    &  1   &  1   &  0    &  174  &  0   &  120  &  360  &  0     &  0\\
2839   &  Annette         &  338  &$-$47  &        &        &  10.46100  &  13  &  17  &  0    &  242  &  13  &  319  &  557  &  0     &  0\\
4231   &  Fireman         &  250  &$-$83  &        &        &  339.57    &  16  &  1   &  26   &  228  &  13  &  248  &  511  &  88    &  0\\
4524   &  Barklajdetolli  &  76   &  70   &  257   &  93    &  966.5     &  19  &  2   &  0    &  175  &  0   &  188  &  391  &  0     &  0\\
5081   &  Sanguin         &  242  &$-$39  &        &        &  10.26457  &  11  &  20  &  41   &  250  &  0   &  248  &  627  &  0     &  0\\
5524   &  Lecacheux       &  254  &$-$86  &        &        &  8.41707   &  5   &  1   &  0    &  196  &  9   &  89   &  468  &  31    &  0\\
5924   &  Teruo           &  184  &$-$60  &  359   &$-$71   &  9.99174   &  6   &  1   &  20   &  153  &  13  &  217  &  447  &  0     &  0\\
9723   &  Binyang         &  36   &  45   &  215   &  38    &  12.38806  &  2   &  1   &  0    &  127  &  0   &  0    &  357  &  0     &  26\\
20771  &  2000 QY150      &  159  &$-$63  &        &        &  8.30140   &  0   &  0   &  0    &  197  &  14  &  26   &  386  &  0     &  0\\
28736  &  2000 GE133      &  241  &$-$65  &        &        &  4.65441   &  3   &  1   &  0    &  237  &  0   &  53   &  493  &  0     &  0\\
\hline
    \end{tabular}
\tablefoot{
Same as in Table~\ref{tab:NewModels}.}
\end{table*}
\end{landscape}

%\onecolumn
%\scriptsize{
%\begin{longtable}{lr rrr l l}
%\caption{\label{tab:lcs}Disk-integrated optical dense light curves utilised for the physical characterisation of primordial family asteroids. }\\
%\hline 
%\multicolumn{1}{c} {Epoch} & \multicolumn{1}{c} {$N_p$} & \multicolumn{1}{c} {$\Delta$} & \multicolumn{1}{c} {$r$} & \multicolumn{1}{c} {$\varphi$} & \multicolumn{1}{c} {Filter} & Reference \\
% &  & (AU) & (AU) & ($^{\circ}$) &  &  \\
%\hline\hline

%*** TABLE B.5 is onlypublished at the CDS.***

%\end{longtable}
%\tablefoot{For each light curve the table gives the epoch,   number of individual measurements $N_p$,   asteroid's distances to the Earth $\Delta$ and the Sun $r$,   phase angle $\varphi$, photometric filter, and   reference to the data.}

\begin{table*}[!ht]
    \renewcommand\thetable{B.6} %As Table B.5 is published only at the CDS.
    \caption{Physical properties of asteroids of the inner main belt primordial family presented in this study.}
    \label{tab:PhysAst}
    \centering
    \begin{tabular}{rlcccccc}
    \hline \hline
    \multicolumn{2}{c}{Asteroid} & $D$ & $\sigma$$D$ & $p_V$ & $\sigma$$p_V$ & Spectral & Ref.\\
    Number & Name/Designation & (km) & (km) &  & & Class &  \\
    \hline
    \multicolumn{8}{c}{Prograde rotators} \\
    \hline
  249 & Ilse            & 31.57   &   0.300   &   0.054   &   0.0014   &   Ch & \cite{Lazzaro2004}  \\
  428 & Monachia        & 20.55   &   0.129   &   0.066   &   0.0055   &  X  &  \cite{alvarez2006} \\
  917 & Lyka            & 35.61   &   0.135   &   0.043   &   0.0040   &  X  & \cite{Lazzaro2004}  \\
 1216 & Askania         & 10.35   &   0.086   &   0.086   &   0.0075   &  -  &   \\
 1544 & Vinterhansenia  & 24.78   &   0.074   &   0.049   &   0.0036   & X,D  &  \cite{carvano2010, alvarez2006} \\ 
 1700 & Zvezdara        & 20.54   &   0.181   &   0.039   &   0.0016   & X   & \cite{zellner1985}  \\
 1806 & Derice          & 10.67   &   0.060   &   0.219   &   0.0512   &  Sl  & \cite{Lazzaro2004}  \\
 2171 & Kiev            & 8.30    &   0.055   &   0.101   &   0.0059   &  S  &  Avdellidou et al. (in prep.) \\
 2536 & Kozyrev         & 9.59    &   0.218   &   0.195   &   0.0227   &  -  &   \\
 2575 & Bulgaria        & 7.92    &   0.063   &   0.270   &   0.0300   & Sr,S  & \cite{Bus2002, popescu2018}  \\
 2768 & Gorky           & 10.67   &   0.085   &   0.258   &   0.0367   & A  & \cite{alvarez2006}   \\
 4024 & Ronan           & 11.90   &   0.073   &   0.055   &   0.0032   & -  &    \\
 4524 & Barklajdetolli  & 12.62   &   0.146   &   0.087   &   0.0069   &  U &  \cite{popescu2018}  \\
 6125 & Singto          & 6.30    &   0.479   &   0.109   &   0.0382   &  S &  \cite{popescu2018}  \\
 8022 & Scottcrossfield & 8.51    &   0.147   &   0.046   &   0.0067   &  C &  \cite{carvano2010}  \\
 8315 & Bajin           & 7.44    &   0.665   &   0.039   &   0.0086   & -  &    \\
 9723 & Binyang         & 3.69    &   0.095   &   0.117   &   0.0320   & -  &    \\
11975 & 1995 FA1        & 4.360   &   1.440   &   0.060   &   0.0500   & -  &    \\
12722 & Petrarca        & 4.52    &   0.746   &   0.095   &   0.0384   &  C,U  & \cite{carvano2010, popescu2018}  \\
30596 & Amdeans         & 5.11    &   0.158   &   0.056   &   0.0074   & -  &    \\
59072 & 1998 VV9        & 4.59    &   0.070   &   0.038   &   0.0045   & -  &    \\
    \hline
    \multicolumn{8}{c}{Retrograde rotators} \\
    \hline
  220 & Stephania       & 32.54   &   0.138   &   0.060   &   0.0032   &   X,Xk & \cite{carvano2010, Lazzaro2004}  \\
  282 & Clorinde        & 34.01   &   0.148   &   0.045   &   0.0019   &   B,C & \cite{Bus2002, popescu2018}  \\
  370 & Modestia        & 38.11   &   0.106   &   0.059   &   0.0046   &   - &   \\
  933 & Susi            & 23.42   &   0.245   &   0.042   &   0.0032   &  C &  \cite{carvano2010, popescu2018}  \\
 1159 & Granada         & 30.14   &   0.099   &   0.046   &   0.0019   &  -  &   \\
 1244 & Deira           & 33.97   &   0.152   &   0.046   &   0.0029   &  X & \cite{Lazzaro2004}  \\
 1705 & Tapio           & 11.81   &   0.061   &   0.093   &   0.0068   &  B,U  &  \cite{Bus2002, popescu2018} \\
 1924 & Horus           & 12.90   &   0.130   &   0.070   &   0.0036   &  -  &   \\
 2012 & Guo Shou-Jing   & 11.91   &   0.076   &   0.048   &   0.0016   &  C  & \cite{carvano2010, popescu2018}  \\
 2322 & Kitt-Peak       & 11.91   &   0.085   &   0.058   &   0.0060   & -   &   \\
 2705 & Wu              & 7.79    &   0.319   &   0.163   &   0.0330   &  -  &   \\
 2772 & Dugan           & 9.58    &   0.135   &   0.057   &   0.0075   &  B  & \cite{Bus2002}  \\
 2773 & Brooks          & 13.37   &   0.100   &   0.042   &   0.0031   &  -  &   \\
 2776 & Baikal          & 13.37   &   0.100   &   0.042   &   0.0031   &  - &    \\
 2778 & Tangshan        & 12.66   &   0.144   &   0.062   &   0.0080   & Cb  &  \cite{Bus2002}  \\
 2792 & Ponomarev       & 12.52   &   0.222   &   0.056   &   0.0112   &  - &    \\
 2839 & Annette         & 7.44    &   0.094   &   0.059   &   0.0076   &  - &    \\
 3633 & Mira            & 10.09   &   0.258   &   0.045   &   0.0058   &  X  &  \cite{carvano2010} \\
 3684 & Berry           & 9.64    &   0.495   &   0.053   &   0.0096   &  C &  \cite{carvano2010, Bus2002} \\
 3723 & Voznesenskij    & 9.54    &   0.029   &   0.041   &   0.0015   &  C &  \cite{carvano2010}  \\
 4231 & Fireman         & 17.96   &   0.557   &   0.022   &   0.0014   & -  &    \\
 5081 & Sanguin         & 17.19   &   0.263   &   0.056   &   0.0063   &  Ch & \cite{Bus2002}   \\
 5333 & Kanaya          & 13.70   &   0.035   &   0.040   &   0.0011   & Ch  &  \cite{Bus2002, morate2019}  \\
 5524 & Lecacheux       & 19.90   &   12.770  &   0.034   &   0.1020   & V  & \cite{carvano2010, popescu2018}   \\
 5924 & Teruo           & 13.16   &   0.080   &   0.059   &   0.0015   &  - &    \\
 6647 & Josse           & 6.42    &   0.148   &   0.049   &   0.0069   &  C  &  \cite{carvano2010} \\
13066 & 1991 PM13       & 8.60    &   0.069   &   0.038   &   0.0058   &  B  & \cite{popescu2018}  \\
15415 & Rika            & 2.83    &   0.488   &   0.605   &   0.1924   &  V &  \cite{popescu2018}  \\
15998 & 1999 AG2        & 7.13    &   0.046   &   0.087   &   0.0062   &  - &    \\
20771 & 2000 QY150      & 9.08    &   0.115   &   0.047   &   0.0068   &  -  &   \\
23495 & 1991 UQ1        & 7.94    &   0.088   &   0.058   &   0.0054   &  -  &   \\
28736 & 2000 GE133      & 7.03    &   0.680   &   0.084   &   0.0186   & C  &  \cite{carvano2010}  \\
49863 & 1999 XK104      & 3.93    &    0.189  &   0.054   &   0.0053   & -  & \\
70184 & 1999 RU3        & 4.31    &   0.629   &   0.137   &   0.0557   & -   &   \\
    \hline
    \end{tabular}
    \tablefoot{$D$ and $p_V$ are the diameter and the geometric visible albedo, respectively. These values are   from the Minor Planet Physical Properties Catalogue and represent the uncertainty-weighted average values of each asteroid. Where the spectral class is known, it is given with its reference.}
\end{table*}

\end{appendix}

\end{document}